\documentclass[lettersize,journal]{IEEEtran}
\usepackage{amsmath,amsfonts}
\usepackage{algorithmic}
\usepackage{algorithm}
\usepackage{array}
\usepackage[caption=false,font=normalsize,labelfont=sf,textfont=sf]{subfig}
\usepackage{textcomp}
\usepackage{stfloats}
\usepackage{url}
\usepackage{verbatim}
\usepackage{graphicx}
\usepackage{cite}
\usepackage{multirow}
\usepackage{color,xcolor}
\usepackage{colortbl}
\usepackage{booktabs}
\usepackage{orcidlink} 
\hypersetup{hidelinks} 

\begin{document}

\title{Mitigating the Impact of Malware Evolution on API Sequence-based Windows Malware Detectors}

\author{Xingyuan Wei \orcidlink{0009-0001-6595-4222},
Ce Li$^{\ast}$ \orcidlink{0009-0000-2405-8854},
Qiujian Lv \orcidlink{0009-0006-2640-4056},
Ning Li \orcidlink{0009-0001-1851-1672},
Degang Sun\orcidlink{0009-0007-6408-2032},
Yan Wang$^{+}$  \orcidlink{0009-0007-7450-4075}\\
This research Accepted by IEEE Transactions on Information Forensics  \& Security in 2025
\thanks{$^{+}$ Corresponding author  \\ $\ast$ Same contribution as the first author}}

\markboth{Journal of \LaTeX\ Class Files,~Vol.~14, No.~8, August~2021}
{Shell \MakeLowercase{\textit{et al.}}: A Sample Article Using IEEEtran.cls for IEEE Journals}

\maketitle
\begin{abstract}
In dynamic Windows malware detection, deep learning models are extensively deployed to analyze API sequences. Methods based on API sequences play a crucial role in malware prevention. However, due to the continuous updates of APIs and the changes in API sequence calls leading to the constant evolution of malware variants, the detection capability of API sequence-based malware detection models significantly diminishes over time. We observe that the API sequences of malware samples before and after evolution usually have similar malicious semantics. Specifically, compared to the original samples, evolved malware samples often use the API sequences of the pre-evolution samples to achieve similar malicious behaviors. For instance, they access similar sensitive system resources and extend new malicious functions based on the original functionalities. In this paper, we propose a framework \textsc{MME}(Mitigating the impact of Malware Evolution), a framework that can enhance existing API sequence-based malware detectors and mitigate the adverse effects of malware evolution. To help detection models capture the similar semantics of these post-evolution API sequences, our framework represents API sequences using API knowledge graphs and system resource encodings and applies contrastive learning to enhance the model's encoder. Results indicate that, compared to regular Text-CNN, our framework can significantly reduce the false positive rate by 13.10\% and improve the F1-Score by 8.47\% on five years of data, achieving the best experimental results. Additionally, evaluations show that our framework can save on the human costs required for model maintenance. We only need 1\% of the budget per month to reduce the false positive rate by 11.16\% and improve the F1-Score by 6.44\%.
\end{abstract}

\begin{IEEEkeywords}
API sequence-based malware detection, Malware evolution, API knowledge graph, Contrastive learning, Deep learning.
\end{IEEEkeywords}

\section{Introduction}
\IEEEPARstart{M}{alware} typically executes its malicious activities through a series of specific system API calls. These API call sequences not only reflect the behavioral patterns of malware but also provide critical dynamic features for detection systems \cite{ChangZK22, DBLP:journals/compsec/UcciAB19, KolosnjajiZWE16, icassp, DBLP:journals/compsec/LiLLWSQ22, DBLP:journals/tifs/ChenHLCZDL22, DBLP:journals/peerj-cs/CatakYEA20, DBLP:conf/acsac/RabadiT20, 2020Amer, zhang2020dynamic}. Using deep neural networks (DNNs) to analyze and identify these API sequences has proven highly effective in dynamic malware detection, as they can capture temporal dependencies and complex patterns within the sequences. However, with the continuous evolution of malware, detection models face challenges such as model aging or concept drift \cite{pendlebury2021machine, DBLP:conf/sp/YangCLA021}, which lead to a significant increase in false positive rates over time \cite{DBLP:conf/uss/PendleburyPJKC19}. Therefore, in-depth research on API sequence analysis to mitigate the impact of malware evolution is crucial for building robust detection systems.

Existing research has proposed various methods to tackle the challenges posed by malware evolution. One common strategy involves retraining models with newly labeled samples through online learning \cite{DBLP:conf/eurosp/XuLDCX19} or active learning \cite{DBLP:conf/uss/PendleburyPJKC19}\cite{Zhang2023SlowingDT}, or rejecting evolved samples for expert analysis \cite{DBLP:conf/uss/Yang0HCAX021, DBLP:conf/uss/JordaneySDWPNC17,MalFSCIL}. However, these approaches demand substantial expert knowledge and computational resources, rendering them costly. An alternative strategy focuses on extending model lifespan by optimizing feature spaces and designing robust models \cite{DBLP:journals/corr/abs-2108-04081, DBLP:conf/asiaccs/DibTBBA22, DBLP:conf/ccs/ZhangZZDCZZY20, DBLP:journals/tdsc/XuLDX22}. For instance, APIGraph \cite{DBLP:conf/ccs/ZhangZZDCZZY20} constructs an API knowledge graph to represent API feature vectors and replaces original inputs with clustered API identifiers, ensuring that APIs with similar semantics share identical cluster identifiers. This method emphasizes semantic similarity between APIs and reduces feature space dimensionality through clustering. Similarly, SDAC \cite{DBLP:journals/tdsc/XuLDX22} calculates semantic distances between APIs based on differences in API vectors within sequences, clustering APIs according to these distances and focusing on their semantic relationships within sequences. While these methods have achieved some success in capturing API-level similarities, they exhibit limitations: they fail to fully exploit system resource access patterns within API parameters and offer limited capability to intervene in existing models. 

\begin{figure}[t]
	\centering
	\includegraphics[scale=0.62]{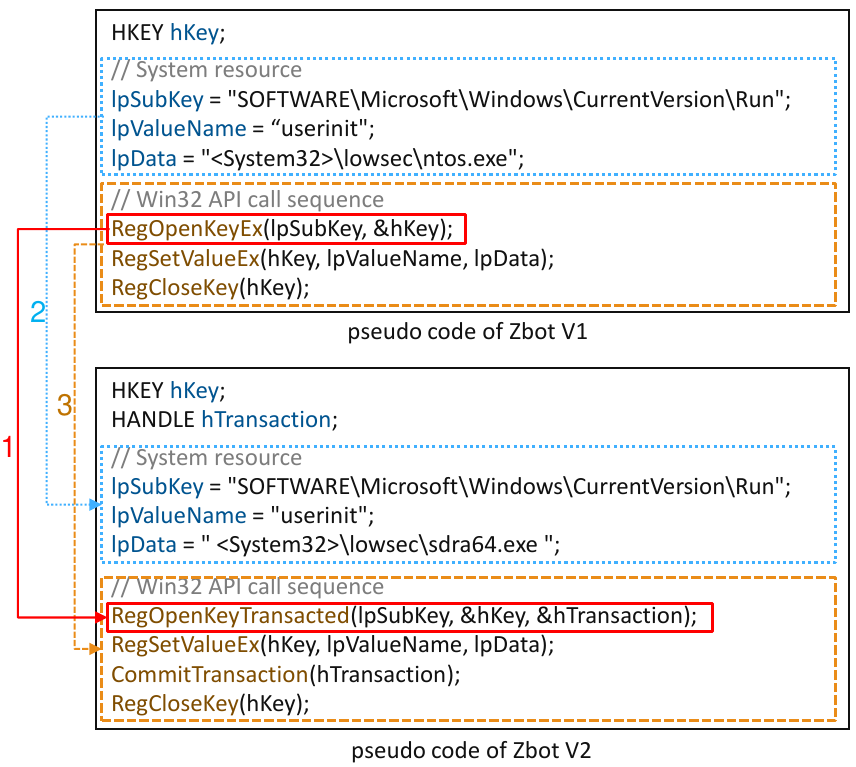}
	\caption{An example to show the similar semantics of API sequences before and after evolution.}
	\label{Motivating example}
\end{figure}

We evaluate and find that the API sequences of malware samples before and after evolution usually have similar malicious semantics. A motivating example about malware \texttt{Zbot} \cite{falliere2009zeus, wyke2011zeus} is shown in Figure \ref{Motivating example}. We reverse two samples (called V1 and V2) and extract the malicious behavior of hiding itself in the registry as a startup entry. This behavior is implemented by calling \texttt{RegOpenKeyEx}, \texttt{RegSetValueEx}, and \texttt{RegCloseKey} in turn, and operating the corresponding system resources (i.e., registry keys and file paths). After evolution, three phenomena can be observed:

\begin{enumerate}
	\item V2 replaces \texttt{RegOpenKeyEx} in V1 with \texttt{RegOpenKeyTransacted}, which means V2 uses transactions to perform the same malicious behavior as V1 for stability. Although the API names of \texttt{RegOpenKeyEx} in V2 and \texttt{RegOpenKeyTransacted} in V1 are different, they represent the same behavior. Intuitively, during evolution, the samples often keep similar behaviors with different implementations using semantically equivalent APIs.
	\item Both V2 and V1 access the similar registry keys (i.e., \texttt{CurrentVersion\textbackslash Run}) and file directories (i.e., \texttt{<System>\textbackslash lowsec}). This indicates that the system resources (such as files, registry keys, URLs, etc) accessed during evolution are highly similar.
	\item V2 still uses some APIs that used in V1 (i.e., \texttt{RegSetValueEx} and \texttt{RegCloseKey}). Actually, during evolution, malware samples often involve massive code reuse and generate similar fragments of the API sequence.
\end{enumerate}





Based on these observations, in this paper, we propose a framework named MME(\textbf{M}itigating the impact of \textbf{M}alware \textbf{E}volution) designed to mitigate model aging and enhance API sequence-based malware detectors from four key perspectives. 

(1) Mining API Parameter Features. MME leverages dynamic analysis to acquire API calls along with their parameters. It employs hierarchical hash encoding on these parameters to extract system resource access patterns (e.g., file paths and registry keys), thereby providing a more comprehensive representation of malware behavior. By fully exploiting the information embedded within API parameters, MME improves the capture of malicious activities. For parameter embedding, system resources accessed by each API operation are extracted from the parameters and represented as fixed feature vectors. These vectors are subsequently concatenated and fed into the detection model.

(2) Enhancing the Model Encoder. MME incorporates contrastive learning to strengthen the model's encoder, boosting its generalization capability against evolving malware. By integrating knowledge graph embeddings with contrastive learning objectives, the encoder measures the similarity between two API sequences by computing the distance between their embedding vectors. Our contrastive learning strategy ensures that malware samples in the feature space are closer to other samples within the same family while maintaining a distinct separation from benign samples. Consequently, as malware undergoes gradual evolution, new samples are expected to resemble past samples (due to shared API sequence fragments), enabling the contrastive encoder to adapt automatically to this evolution. This approach enhances the encoder’s ability to capture both semantic and temporal robustness.

(3) Constructing a Knowledge Graph for Richer Semantic Representation. MME constructs a knowledge graph using API documentation \cite{Win32API} to capture semantic relationships between APIs. This construction process is more thorough, integrating API function descriptions and parameter information to deliver a richer semantic representation. As a result, MME effectively identifies similarities between APIs, enhancing the overall understanding of their behavioral context. 

(4) Augmenting Existing Models. MME is designed to enhance existing API sequence-based models, improving their robustness in dynamic environments. For instance, it can be seamlessly integrated with models such as LSTM \cite{staudemeyer2019understanding}, Text-CNN \cite{DBLP:conf/emnlp/Kim14}, Transformer \cite{transformer}, and BERT \cite{bert}, further elevating their performance. This adaptability makes MME a versatile solution for advancing malware detection capabilities. 

To evaluate our approach, MME is used to enhance two classic API sequence detection models, namely long short-term memory networks (LSTM) and text convolutional neural networks (Text-CNN), as many work  use their variants or combinations as detection models \cite{KolosnjajiZWE16, icassp, zhang2020dynamic, DBLP:journals/peerj-cs/CatakYEA20, DBLP:journals/compsec/LiLLWSQ22, DBLP:journals/tifs/ChenHLCZDL22}. We collect about 76K Windows PE samples spanning from 2017 to 2021. We train the regular models and enhanced models using data in 2017 and evaluate their performance of them from 2018 to 2021. Our evaluation shows that MME can significantly mitigate the model aging of the malware detectors. It reduces the average false negative rate from 22.4\% to 10.1\% for LSTM, and from 22.7\% to 9.6\% for TextCNN. Additionly, MME can significantly reduce the amount of human analyst effort required for model periodical retraining and maintenance. The number of samples needed to be labeled can be reduced by 24.19\%-94.42\%. Finally, model ablation analysis and feature stability analysis explore why MME can help the model mitigate the impact of malware evolution.

To summarize, we make the following contributions in this paper:
\begin{itemize}
	\item We first observe that the API sequences of malware samples before and after evolution usually have similar malicious semantics, including equivalent APIs, similar system resources, and similar API fragments. This provides an opportunity to reduce the feature gaps caused by evolution, and to slow down model aging (\S I).
    
	\item We design a framework called MME to enhance the API sequence-based malware detectors (\S II). MME contains a new API embedding method to capture the similarities between APIs (\S III and \S IV), and a contrastive learning strategy to enhance the encoder of the detection model (\S V).
    
	\item We apply MME to two widely used Windows malware detection models. The results show that MME can significantly reduce the high false negative rate caused by malware evolution, thereby slowing down model aging. Thus, the aging speed of the model is slowed down. At the same time, MME can also significantly save the workload of manual labeling when retraining the model, showing its universality and the double optimization of reducing labor costs. (\S VII).
\end{itemize}

\section{Design Overview}
\begin{figure*}[t]
	\centering
	\includegraphics[width=\textwidth]{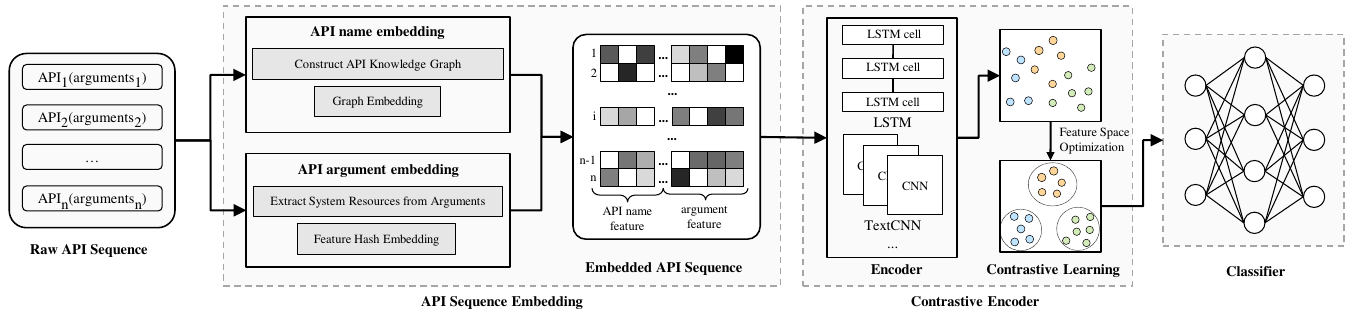}
	\caption{Framework Overview of MME. MME focuses on enhancing the API sequence embedding and encoder modules.}
	\label{system_overview}
        \vspace{-1.0em}
\end{figure*}

Figure \ref{system_overview} shows the overview architecture of MME. Generally, a DNN malware detection model consists of three parts: API sequence embedding, encoder, and classifier. First, API sequence embedding represents each raw API sequence as feature vectors (i.e., embedded API sequence) and inputs them to the encoder. Then, the encoder learns the features and maps each embedded API sequence to the feature space. Finally, the classifier learns the samples in the feature space and outputs the prediction results (i.e., malware or goodware).

Our framework MME focuses on enhancing the API sequence embedding and encoder modules. For API sequence embedding enhancement, we first construct an API knowledge graph that can find semantically equivalent APIs and use graph embedding to represent API names (\S III). Then, to capture the system resources operated by each API, we use feature hash embedding to represent the arguments of each API (\S IV). For encoder enhancement, we design a contrastive learning strategy to help the model learn the similarities of samples in the same malware family, while learning the dissimilarity between malware and goodware (\S V). Finally, the enhancement can be achieved simply by adding MME's API sequence embedding and contrastive learning strategy to the original model, without altering the original model structure.

\begin{figure}[t]
	\centering
	\includegraphics[scale=0.65]{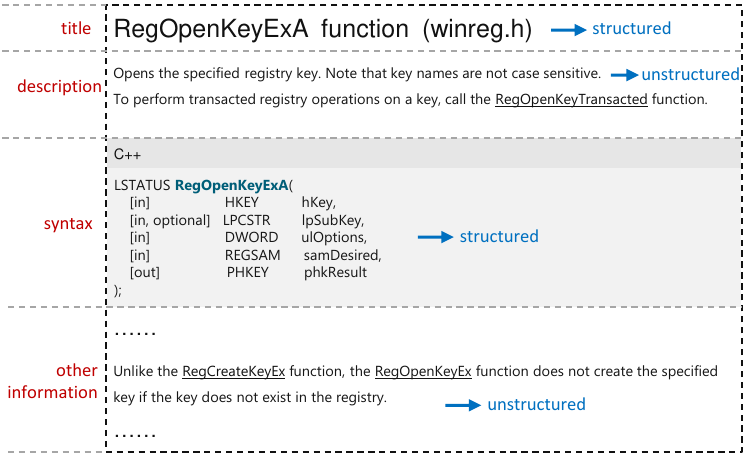}
	\caption{The API documentation for RegOpenKeyEx.}
	\label{API document example}
        \vspace{-1.0em}
\end{figure}

\section{API Name Embedding}\label{sec:API Name Embedding}
In this section, we analyze the Windows API documents \cite{Win32API} and construct a API knowledge graph, which can capture the similarities between APIs and represent API names as semantic feature vectors. 

\begin{figure}[t]
	\centering
	\includegraphics[scale=0.62]{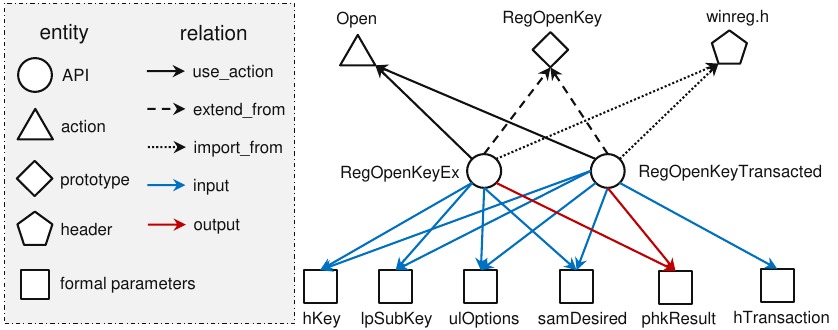}
	\caption{An example to show API knowledge graph.}
	\label{API KG example}
        \vspace{-1.0em}
\end{figure}

We first explain how the knowledge graph can capture the semantic similarity between APIs. The components of API documentation, using \texttt{RegOpenKeyEx} as an example, are shown in Figure \ref{API document example}. Some API-related entities can be extracted from this document, such as action \texttt{Open} (mentioned in the first sentence of the description), prototype \texttt{RegOpenKey} (remove the suffix of the API name), header \texttt{winreg.h} (mentioned in the title), and formal parameters (mentioned in the syntax). Figure \ref{API KG example} shows a small part of the knowledge graph, which captures the relations between the equivalent APIs of \texttt{RegOpenKeyEx} and \texttt{RegOpenKeyTransacted}. Intuitively, these two APIs use the same action, extend from the same prototype, and import from the same header. Besides, they have very similar input/output parameters. That is, these APIs are similar enough in terms of their neighborhoods in the graph. If two APIs are connected to more identical entities, their semantics will become more similar. Therefore, the knowledge graph can capture the similarity between equivalent APIs and then help detectors to detect evolved malware.

In the next subsections, we will introduce the API knowledge graph construction (\S\ref{sec:API Knowledge Graph Construction}) and use graph embedding to represent API names as semantic feature vectors (\S\ref{sec:Graph Embedding}).

\subsection{API Knowledge Graph Construction}\label{sec:API Knowledge Graph Construction}
\subsubsection{API Documents Collection}

To construct the knowledge graph, the Windows API documents are collected. As shown in Figure \ref{API document example}, each document consists of four parts: \textit{title}, \textit{description}, \textit{syntax} and \textit{other information}. Among these four parts, \textit{title} and \textit{syntax} are structured texts, which contain the basic information of the API (i.e., API name, source header file, class to which it belongs, and function declaration). The \textit{description} and \textit{other information} are unstructured texts that contain specific descriptions of API functions and the relationship between the current API and other APIs. We downloaded the API documents for Windows 10 from the official website \cite{Win32API} and analyzed them to construct an API knowledge graph.

\subsubsection{Knowledge Graph Construction}
The API knowledge graph $G = \langle E, R \rangle$ is defined as a directed graph, where $E$ is the set of all nodes (called entities), and $R$ is the set of all edges (called relations) between two nodes. API knowledge graph is heterogeneous, which means that entities and relations have different types.

There are six types of entities and eight types of relations extracted from API documents to construct the API knowledge graph. Table \ref{Entities} lists the specific entities of the graph. For entity extraction, we first consider four basic concepts in Windows API documentation: \textit{API}, \textit{header}, \textit{class}, and \textit{parameter}. These four entities can be extracted directly from the API documentation. Specifically, \textit{API}, \textit{header}, and \textit{class} can be extracted from the \textit{title}. The input and output \textit{parameter}s can be extracted from the \textit{syntax}. Using function \texttt{RegOpenKeyEx} in Figure \ref{API document example} as an example, the entity \textit{API} is \texttt{RegOpenKeyEx} and \textit{header} is \texttt{winreg.h}, which can be extracted from the \textit{title}. Several \textit{parameter}s (including input parameters \texttt{hKey}, \texttt{lpSubKey}, etc., and an output parameter \texttt{phkRFesult}) are extracted from the \textit{syntax}. Then, the other two types of entities, namely \textit{action} and \textit{prototype}, can be extracted after analyzing the content of the API documentation. Specifically, for each API document, the first sentense of the \textit{description} is a summary, where the verbs are extracted as the \textit{action} of the API. The \textit{action} can reflect the semantic similarity between APIs. For example, the \textit{action}s of \texttt{RegOpenKeyEx} and \texttt{RegOpenKeyTransacted} are both \texttt{Open}, and the \textit{action}s of \texttt{GetFileSize} and \texttt{GetFileType} are both \texttt{Retrieve}. Another type of entity that reflects semantic similarity is \textit{prototype}. We found that many similar APIs are extended from the same prototype by adding various suffixes in order to adapt to different system environments, but their functions have not changed. For example, the APIs of \texttt{RegOpenKeyA/W}, \texttt{RegOpenKeyExA/W}, \texttt{RegOpenKeyTransactedA/W} are extended from the \textit{prototype} of \texttt{RegOpenKey}. Thus, for each API name, we remove some specific suffixes (including \texttt{A}, \texttt{W}, \texttt{Ex}, \texttt{Transacted}, \texttt{Advanced}, and \texttt{0-9}) and get its \textit{prototype}.

\begin{table}[t]
	\renewcommand\arraystretch{1}
	\caption{Entities}
	\label{Entities}
	\tiny
	\resizebox{\linewidth}{!}{
		\begin{tabular}{llll}
			\toprule
			Entity Type & Examples  & Related Source   & Count \\
			\midrule
			API  & RegOpenKeyExA, CreateFileA & title & 40,472             \\
			header   & fileapi.h, winbase.h & title &  795 \\
			class    & IUnknown  & title           & 4,242 \\
			parameter  & hKey, pSubKey  & syntax   & 27,438 \\
			action   & Open, Create, Write  & description      & 756 \\
			prototype  & RegOpenKey, CreateFile  & title            & 3,163 \\
			\bottomrule
		\end{tabular}
	}
  \vspace{-1.0em}
\end{table} 

For relations, we extracted a total of eight relations (as shown in Table \ref{Relations}). Among them, six types of relations can be directly established after entity extraction:
\begin{itemize}
	\item \textit{function\_of}: connects an \textit{API} to its belonging \textit{header} or \textit{class}.
	\item \textit{inheritance}: connects a \textit{class} entity with its inherited \textit{class} entity. It can be extract from the ``Inheritance" section in the \textit{class} definition document. The sentence template is ``the \textit{class} inherits from the \textit{class}".
	\item \textit{input}: connects an \textit{API} to its input \textit{parameter}.
	\item \textit{output}: connects an \textit{API} to its output \textit{parameter}.
	\item \textit{use\_action}: connects an \textit{API} to its \textit{action}.
	\item \textit{extend\_from}: connects an \textit{API} to its \textit{prototype}.
\end{itemize}

Furthermore, the remaining two types of relations, namely \textit{bundled\_with} and \textit{replaced\_by}, are used to describe the relationships between APIs. The \textit{bundled\_with} refers to the relation that two APIs must be used at the same time, such as a program must call \texttt{DestroyWindow} once for every time it called \texttt{CreateWindow}. The \textit{replaced\_by} means that the two APIs are functionally equivalent and can be used instead, such as \texttt{RegOpenKeyEx} and \texttt{RegOpenKeyTransacted}. These two types of relations can be derived from the unstructured text within the API documentation. However, manually extracting these relations one by one from unstructured text is impractical due to the large number of \textit{API} documents involved. We have observed that there are common patterns when describing the relations between \textit{API} entities. These patterns can be summarized with templates and used them for relation extraction. The template-based relation extraction involves three steps. Firstly, for all API documents, we employ NLP tools to tokenize each unstructured text into sentences and normalize the sentences. Secondly, we select sentences that contain more than one \textit{API} entity to form a corpus. Thirdly, we employ a semi-automated strategy to analyze the sentences in the corpus and iteratively formulate templates for relation matching. Table \ref{two types of relations} provides several example templates in regular expression format for relations of \textit{bundled\_with} and \textit{replaced\_by}. The detailed process is as follows:

i) \textit{Sentence tokenization and normalization.} For each API document, we use spaCy \cite{spaCy} (a Python NLP toolkit) for text processing. We first split the unstructured text into sentences. For each sentence, we check if it is a sentence lacking a subject. If it is, we supplement the subject of the sentence with the corresponding API entity it describes. Then, we employ the coreference resolution \cite{NeuralCoref} to convert pronouns in the sentence into their corresponding entities.

ii) \textit{Sentence selection.} We employ named entity recognition to extract entities from each sentence, and select sentences that contain more than one \textit{API} entity to form a corpus. After this step, the scale of the data we need to analyze has been reduced from about 40K API documents to 10K sentences in the corpus.

iii) \textit{Template iteratively generation}. For each sentence in the corpus, we manually check whether there is \textit{bundled\_with} or \textit{replaced\_by} relation between two \textit{API} entities. If the answer is no, we remove that sentence from the corpus. Otherwise, we manually formulate a template for the relation and use it for regular expression matching with all the sentences in the corpus. For the sentences that match this template, we extract the corresponding relation from the sentence and remove the sentence from the corpus. Finally, we repeat this process until there are no sentences in the corpus.

In total, 76,886 entities and 215,467 relations are extracted to build the API knowledge graph. If two APIs are connected to more identical entities, their semantics will become more similar. Next, we will use graph embedding to represent API names as semantic feature vectors.

\begin{table}[t]
	\renewcommand\arraystretch{1}
	\caption{Relations}
	\label{Relations}
	\Large
	\resizebox{\linewidth}{!}{
	\begin{tabular}{lllll}
		\toprule
		Relation Type & Entity Connection                                 & Examples                                        & Related Source   & Count \\
		\midrule
		function\_of  & \begin{tabular}[c]{@{}l@{}} API $\xrightarrow{}$ header,\\ API $\xrightarrow{}$ class\end{tabular} & RegOpenKeyExA $\xrightarrow{}$ winreg.h             & title & 64,217             \\
		inheritance   & class $\xrightarrow{}$ class                         & Istream $\xrightarrow{}$ Isequentialstream         & unstructured text &  3,501 \\
		input         & API $\xrightarrow{}$ parameter                       & RegOpenKeyExA $\xrightarrow{}$ hKey                 & syntax           & 76,967 \\
		output        & API $\xrightarrow{}$ parameter                       & RegOpenKeyExA $\xrightarrow{}$ phkResult            & syntax           & 22,834 \\
		use\_action   & API $\xrightarrow{}$ action                          & RegOpenKeyExA $\xrightarrow{}$ Open                 & description      & 38,683 \\
		extend\_from  & API $\xrightarrow{}$ prototype                       & RegOpenKeyExA $\xrightarrow{}$ RegOpenKey           & title            & 6,060 \\
		bundled\_with & API $\xrightarrow{}$ API                             & CreateWindow $\xrightarrow{}$ DestroyWindow        & unstructured text & 421 \\
		replaced\_by  & API $\xrightarrow{}$ API                             & RegOpenKeyEx $\xrightarrow{}$ RegOpenKeyTransacted & unstructured text & 2,784 \\
		\bottomrule
	\end{tabular}
}
\end{table}

\begin{table}[t]
	\renewcommand\arraystretch{1.2}
	\caption{Templates to extract ralations of bundled\_with and replaced\_by}
	\label{two types of relations}
	\resizebox{\linewidth}{!}{
	\begin{tabular}{llc}
		\toprule
		Relation      & Example Templates                                                                                                                                                                                                                                                                                                                                                                                                                                 & \# of Templates \\ \midrule
		bundled\_with & \begin{tabular}[c]{@{}l@{}}call \textit{API} once for every time it called \textit{API}\\ for every successful call to \textit{API}, there should be a ... call to \textit{API}\\ \textit{API} must be called at the same depth at which \textit{API} was called\\ call \textit{API} before calling \textit{API}\end{tabular} & 58              \\ \midrule
		replaced\_by  & \begin{tabular}[c]{@{}l@{}}To perform ..., call \textit{API}\\ \textit{API} is superseded by the \textit{API}\\ not necessary to call \textit{API} when \textit{API} is called\end{tabular}                                                                                                                                                                                 & 27              \\ \bottomrule
	\end{tabular}
}
\end{table}

\subsection{Graph Embedding}\label{sec:Graph Embedding}
Graph embedding \cite{DBLP:conf/nips/BordesUGWY13, DBLP:conf/aaai/WangZFC14, DBLP:conf/aaai/LinLSLZ15} can represent each \textit{API} in the knowledge graph as a feature vector. Moreover, the semantically similar APIs are represented closer in the feature space.
To achieve this, we employed an existing algorithm called TransE \cite{DBLP:conf/nips/BordesUGWY13} and integrated it into our graph embedding problem. Specifically, suppose there is a relation $R$ that connects the entity $E_a$ to the entity $E_b$, and they are represented by three vectors: $V_R$, $V_a$, and $V_b$. The core idea of the TransE algorithm is to iteratively adjust these three vectors so that the $V_a + V_R$ is as close as possible to $V_b$. As a result, APIs with similar semantics will have similar vector representations because they will be related to the same other entities. As a result, the entities with similar semantics will have similar vector representations in the vector space. For example, the two API entities \texttt{RegOpenKeyEx} (denoted as $V_{a1}$) and \texttt{RegOpenKeyTransacted} (denoted as $V_{a2}$) have the same prototype \texttt{RegOpenKey} (denoted as $V_b$). Thus, there are \textit{extend\_from} relations (denoted as $V_R$) connect $V_{a1}$ and $V_{a2}$ to the $V_b$. TransE adjusts these vectors so that the $V_{a1} + V_R$ and the $V_{a2} + V_R$ are as close as possible to $V_b$. Therefore, the $V_{a1}$ and $V_{a2}$ are represented more similar.

After graph embedding, each \textit{API} entity in the API knowledge graph is represented as a fixed-length semantic vector. In other words, for the input of the raw API sequence, the API name of each API can be mapped to the corresponding semantic vector using the knowledge graph. Furthermore, when malware undergoes API replacement during its evolution, even though the API names before and after evolution may differ, if API functions are similar, then their semantic vectors will be very close.

\section{API Argument Embedding}\label{sec:API Argument Embedding}
Based on our observations, malware tends to access similar system resources (such as files, registry keys, etc.) before and after evolution. These accessed resources can be extracted from the hooked API sequences during the software execution. Each API call in the sequence consists of two parts: the API name and the arguments. In this section, we extract the system resources accessed during software execution from the API arguments and represent them as semantic feature vectors. This allows detection models to capture the semantic similarity of samples before and after evolution.

\subsection{Extract System Resources from Arguments}
\begin{figure}[t]
	\centering
	\includegraphics[width=0.8\linewidth]{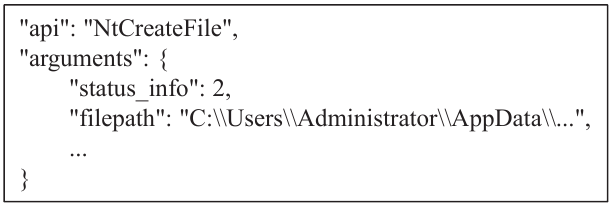}
	\caption{One example hooked API in the API sequence.}
	\label{API example}
\end{figure}

Figure \ref{API example} shows an example hooked API whose name is \texttt{NtCreateFile}. For the first argument, its type is integer, and the value is 2. For the second argument, its type is string and the value is ``\texttt{C:$\backslash\backslash$User$\backslash\backslash$Administrator$\backslash\backslash$AppData$\backslash\backslash$...}" which is an accessed file path. 

To extract system resources, we consider 5 types of string arguments: file paths, dynamic link library file names (DLLs), registry keys, URLs, and IP addresses. These types of resources are accessed frequently and can be extracted directly from the API sequence. For each API in the API sequence, we use regular expression matching to identify its argument values and extract arguments belonging to the 5 types of resources. Specifically, we use ``\texttt{C:$\backslash\backslash$}" to identify a file path. The DLLs are arguments ending with ``\texttt{.dll}". The registry keys often start with ``\texttt{HKEY\_}". URLs often start with ``\texttt{http}". IPs are those arguments with four numbers (range from 0 to 255) separated by dots. These extracted string arguments are then embedded as feature vectors.

\subsection{Feature Hash Embedding}
Intuitively, the strings sharing a large number of substrings have very similar meanings. Thus, for each extracted argument, we first parse the whole string into several substrings to capture the hierarchical information. For example, for a path like ``\texttt{C:$\backslash\backslash$f\_a$\backslash\backslash$f\_b}", three substrings are generated by splitting based on ``\texttt{$\backslash\backslash$}", namely ``\texttt{C:}", ``\texttt{C:$\backslash\backslash$f\_a}", ``\texttt{C:$\backslash\backslash$f\_a$\backslash\backslash$f\_b}". The DLLs and registry keys can also be parsed like the file paths. The URLs and IP address can be parsed by splitting based on ``\texttt{.}". For example, for a URL ``\texttt{https://sample.sec.org/}", we only generate substrings from its hostname, and the following substrings will be generated ``\texttt{org}", ``\texttt{sec.org}", and ``\texttt{sample.sec.org}".

We use feature hashing \cite{DBLP:conf/icml/WeinbergerDLSA09} to represent each extracted argument as a fixed-length feature vector. Let $S$ denotes a substring set of the extracted string argument, and $s_j \in S$ denotes a substring. Let $N$ denote the number of bins. The value of the $i$-th bin is calculated by
\begin{equation}\label{Hash}
	\phi_{i}^{h, \xi}(S)=\sum_{j:h\left(s_{j}\right)=i}\xi\left(s_{j}\right),
\end{equation}
where $h$ is a hash function that maps the $s_j$ to a natural number $n_1 \in \{1, 2, . . ., N\}$ as the bin index. $\xi$ is another hash function that maps the $s_j$ to $n_2 \in \{\pm 1\}$. After feature hashing, the extracted argument $S$ is represented as a feature vector $[\phi_{1}^{h, \xi}(S), \phi_{2}^{h, \xi}(S),..., \phi_{N}^{h, \xi}(S)] \in \mathbb{R}^N$. For example, for the url ``\texttt{https://sample.sec.org/}", if $N$ = 8 and $S$ = \{``\texttt{org}", ``\texttt{sec.org}", ``\texttt{sample.sec.org}"\}, then $s_1$ = ``\texttt{org}", $s_2$ = ``\texttt{sec.org}", $s_3$ = ``\texttt{sample.sec.org}". After hash mapping, $h\left(s_{1}\right)$ = 1 (i.e., bin index 1), $\xi\left(s_{1}\right)$ = 1, $h\left(s_{2}\right)$ = 2 (i.e., bin index 2), $\xi\left(s_{2}\right)$ = -1, $h\left(s_{3}\right)$ = 4 (i.e., bin index 4), $\xi\left(s_{3}\right)$ = 1. Thus, $\phi_{1}^{h, \xi}(S)$ = 1, $\phi_{2}^{h, \xi}(S)$ = -1, $\phi_{4}^{h, \xi}(S)$ = 1. The feature vector of the extracted argument is $[1, -1, 0, 1, 0, 0, 0, 0]$. 

The arguments with a large number of shared substrings will have the similar set $S$ and will be represented very similarly. In this way, if the malware accesses similar system resources before and after evolution, then their feature vectors will be very close.

At this point, the API sequence embedding enhancement is complete. When a raw API sequence is input to the model, for each API in the sequence, its API name is mapped to the API knowledge graph and represented as an API name semantic vector. Each argument of the API is checked to identify if it is an accessed resource. If so, it is hashed and represented as an API argument semantic vector. These two vectors are concatenated as the API's feature vector. Finally, the embedded API sequence (i.e., the API feature vector sequence) is input to the encoder of the detection model.

\section{Contrastive Encoder}
Based on our observations, during evolution, malware samples often involve massive code reuse and generate similar API sequence fragments. In this section, we enhance the encoder's attention to similar API sequence fragments by designing a contrastive learning strategy. Through contrastive learning, the encoder can measure the similarity of two embedded API sequences by calculating the distance between them, and make malware closer to samples with similar API fragments and farther away from benign samples in the feature space. Thus, when a malware sample experiences gradual evolution, it can be expected that the representation of new samples will be similar to past samples and the contrastive encoder can automatically adapt to evolution. Our approach focuses on family-level similarity and seeks to ensure that malware variants evolving within the same family are closely embedded in the feature space, thus mitigating model aging caused by API substitutions and sequence variations.

\begin{figure}[t]
	\centering
	\includegraphics[scale=0.65]{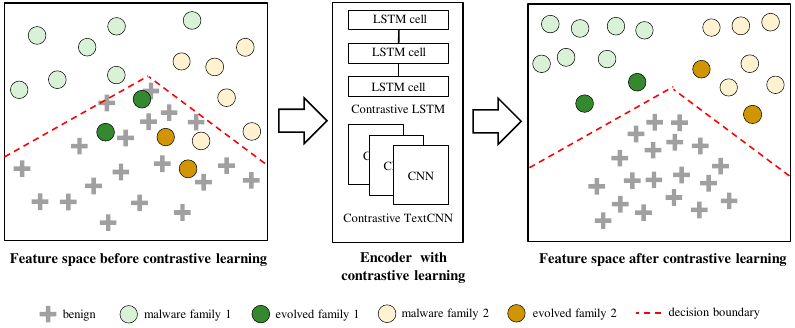}
	\caption{The high-level idea of contrastive learning.}
	\label{CL example}
        \vspace{-1.0em}
\end{figure}

As shown in Figure \ref{CL example}, given the input samples with feature vectors, the contrastive learning encoder aims to map them into a latent feature space. Before contrastive learning, the evolved malware produces many differences in the feature space, leading the detection model to misclassify it as benign. Then, the contrastive learning optimizes the encoder and generates a latent space. In the latent space, pairs of samples in the same class have a smaller distance, and pairs of samples from different classes have a larger distance. As such, the encoder will pay more attention to the similarities among samples from the same malware family. Any evolved sample that retains similar API fragments to the past samples will be represented as closer, thereby reducing misclassifications of the evolved samples.

\subsection{Contrastive Learning Strategy}
We design a contrastive learning strategy to enhance the encoder's ability to capture fine-grained similarities and differences among API sequences and improve performance in detecting evolved malicious samples.

Let $x$ be an embedded API sequence. The ground truth binary label is $y \in \{0, 1\}$, where $y = 0$ indicates a benign sample, and $y = 1$ indicates a malicious sample. Let $y\prime$ be the ground truth multi-class family label. When $y\prime = 0$, the label is benign, but otherwise, it is a malware family label. For the detection model $f$, after API sequence embedding, the embedded sample $x$ is first input to an encoder $en$ (e.g., LSTM, Text-CNN, etc.), which outputs the representation of the input sample in the latent feature space $z = en(x)$. Then, a classifier $g$ takes the encoder output and predicts the binary label $f(x) = g(z) = g(en(x))$.

Let $f(x) = g(en(x))$ be the output of the softmax layer for class $y = 1$ (i.e., malware) and the benign softmax output is $1 - f(x)$. If $f(x) \geq 0.5$, the predicted binary label $\hat{y}$ is $\hat{y} = 1$, and otherwise, $\hat{y} = 0$.

In general, the training loss of a regular model is defined as computing a classification loss between $f(x)$ and $y$. However, in this paper, we define the training loss is the sum of a contrastive loss and a classification loss, and the detection models are trained end-to-end with this loss. Specifically,

\begin{equation}
	\mathcal{L}=\mathcal{L}_{con}+\lambda \mathcal{L}_{cla}
\end{equation}

where $\mathcal{L}_{cla}$ is the classification loss and $\mathcal{L}_{con}$ is the contrastive loss for enhancing the encoder (defined below). As a common heuristic approach, we use a hyperparameter $\lambda$ to balance the two terms $\mathcal{L}_{con}$ and $\lambda \mathcal{L}_{cla}$, so that they have a similar mean, thus the overall loss is not overwhelmed by just one term. The classification loss $\mathcal{L}_{cla}$ uses the binary cross entropy loss:
\begin{equation}
	\begin{gathered}
		\mathcal{L}_{cla}=\sum_i \mathcal{L}_{cla}\left(x_i, y_i\right) \\
		\mathcal{L}_{cla}\left(x_i, y_i\right)=-y_i \log f\left(x_i\right)-\left(1-y_i\right) \log \left(1-f\left(x_i\right)\right)
	\end{gathered}
\end{equation}
where $i$ ranges over indices of samples in the batch.

The contrastive loss $\mathcal{L}_{con}$ computes a similarity over positive and negative pairs of samples in a batch. It tends to maximize the similarity between positive pairs and minimize the similarity between negative pairs. We design a contrastive learning strategy that encourages the encoder $en$ to satisfy the following two properties:

\begin{itemize}
	\item \textit{Positive pairs}: If $x_1$, $x_2$ are two benign samples, or two malicious samples in the same malware family, then they are positive pairs, and their representations should be similar: i.e., $\left\|e n\left(x_1\right)-e n\left(x_2\right)\right\|_2$ should be as small as possible.
	\item \textit{Negative pairs}: If one of $x_1$, $x_2$ is malicious and the other is benign, then they are negative pairs, and their representations should be dissimilar: i.e., $\left\|e n\left(x_1\right)-e n\left(x_2\right)\right\|_2$ should be as large as possible.
\end{itemize}

Specifically, for a batch of size $2N$, the first $N$ samples in the batch are sampled randomly, denoted as $\{x_k, y_k, y\prime_k\}_{k=1...N}$. Then, we randomly select $N$ more samples which have the same label distribution as the first $N$ samples, i.e., $\{x_{k+N}, y_{k+N}, y\prime_{k+N}\}_{k=1...N}$ are chosen so that $y_k = y_{k+N}$ and $y\prime_k = y\prime_{k+N}$. To capture the positive and negative samples paired with $x_i$, the following sets are defined in the batch:
\begin{itemize}
	\item The positive sample set of $x_i$. Both samples are benign or both samples are malicious and in the same malware family:
	\par $Pos\left(x_i\right) \equiv\left\{x_j \mid y_j = y_i, y_i = 1 \Longrightarrow y_j^{\prime} = y_i^{\prime}, j \neq i\right\}$ 
	\item The negative sample set of $x_i$. One sample is benign and the other is malicious:
	\par $Neg\left(x_i\right) \equiv\left\{x_j \mid y_j \neq y_i, j \neq i\right\}$
\end{itemize}
Intuitively, $Pos\left(x_i\right)$ contains samples that are considered similar to $x_i$, and $Neg\left(x_i\right)$ contains dissimilar samples to $x_i$.

Let $d_{i j}$ denote the euclidean distance between two arbitrary samples $x_i$ and $x_j$ in the feature space: $d_{i j} = \left\|e n\left(x_1\right)-e n\left(x_2\right)\right\|_2$. Let $m$ denote a fixed margin (a hyperparameter). The contrastive loss is defined as:
\begin{equation}
	\mathcal{L}_{con}=\sum_{x_i \in Batch} \mathcal{L}_{con}(x_i)
\end{equation}
\begin{equation}
	\begin{aligned}
		\mathcal{L}_{con}(x_i) & =\frac{1}{\left|Pos\left(x_i\right)\right|} \sum_{x_j \in Pos\left(x_i\right)} d_{i j} \\
		& +\frac{1}{\left|Neg\left(x_i\right)\right|} \sum_{x_j \in Neg\left(x_i\right)} \max \left(0, m-d_{i j}\right)
	\end{aligned}
\end{equation}

The contrastive loss has two terms. The first term asks positive pairs from $Pos\left(x_i\right)$ to be close together. These pairs are (benign, benign) or (malicious, malicious) pairs with the same malware family. In this way, the encoder will pay more attention to the similarities among samples in the same class. The evolved samples that retain API fragments similar to past samples will be represented as closer to past samples in the latent space. The second term aims to separate benign and malicious samples from each other, hopefully at least $m$ apart from each other. Thus, the encoder will focus on capturing the differences between benign and malicious samples, and prevent the classifier from misclassifying evolved malware as benign samples.

At this point, the encoder enhancement is complete. A contrastive encoder is constructed using our contrastive learning strategy, without altering the structure of the original model. Finally, the enhanced models are trained end-to-end with the loss $\mathcal{L}$.

\section{Experimental Setup}
In this section, we describe the datasets and baseline malware detectors used in our experiments.

\subsection{Dataset}
In this paper, we focus on malware of the Windows portable executable (PE) file which is the most popular malware file format. Our dataset, spanning over five years, contains 76,473 Windows PE files, i.e., 39,349 malicious and 37,124 benign as shown in Table \ref{Dataset}. Specifically, the malicious software is obtained from the VirusShare website \cite{VirusShare} and using a daily downloading script. The benign software is obtained from popular free software sources, including PortableApps \cite{Portableapps}, Softonic \cite{Softonic}, SourceForge \cite{Sourceforge}, and CNET \cite{CNET}, BODMAS\cite{BODMAS}. We made some of the data files publicly available\footnote{https://github.com/XingYuanWei/MME}. 

To get reliable labels for these samples, we rely on VirusTotal\cite{VirusTotal} to determine whether a sample is benign or malicious. VirusTotal uses more than 60 anti-virus (AV) engines to vote whether the submitted sample is malicious or benign. In this paper, samples are labeled as malware when at least 10 AV engines report them as malicious, while samples are labeled as benign when no AV engines report them as malicious. Note that according to a recent study \cite{DBLP:conf/uss/ZhuSYQZS020} on measuring the labeling effectiveness of malware samples, this strategy is reasonable and stable. We consider samples up to Dec 2021 because following a previous work \cite{DBLP:conf/dimva/MillerKTABFHSWY16}, the malware labels become stable after about one year, thus choosing Dec 2021 as the finishing time ensures good ground-truth confidence in objects labeled as malware.

Also, we leverage VirusTotal \cite{VirusTotal} to get the exact appearing time for each sample and make sure that temporal consistency \cite{DBLP:conf/uss/PendleburyPJKC19} is satisfied at the month level during the testing. Specifically, temporal consistency ensures that training samples should be strictly temporally precedent to testing ones, and all testing samples must come from the same period during each testing to eliminate time bias.

\begin{table}[]
        \tiny
	\renewcommand\arraystretch{1}
	\caption{Dataset}
	\label{Dataset}
	\resizebox{\linewidth}{!}{
		\begin{tabular}{lcccccc}
			\toprule
			Year & 2017 & 2018  & 2019  & 2020  & 2021  & Total \\
			\midrule
			Goodware & 5,788 & 6,748 & 9,976 & 5,961  & 8,651 & 37,124 \\
			Malware  & 3,517 & 6,130  & 7,557 & 9,556 & 12,589 & 39,349 \\ 
			Total    & 9,305 & 12,878  & 17,533 & 15,517 & 21,240 & 76,473 \\ 
			\bottomrule
		\end{tabular}
	}
\end{table}

\subsection{API Sequence Extraction}
After data collection, the Cuckoo Sandbox \cite{Cuckoo} is used to run the PE files and gather execution logs. Cuckoo sandbox has been widely used in prior works \cite{DBLP:journals/peerj-cs/CatakYEA20, DBLP:conf/acsac/RabadiT20, 2020Amer, zhang2020dynamic}. It executes each PE file inside virtual machines and uses API hooks to monitor the Windows APIs to form a raw API sequence. In our system, dozens of virtual machines are maintained on the Cuckoo server which is installed with Ubuntu 16.04 LTS. All the virtual machines are installed with a 64-bit Windows 10 system and several necessary drivers to ensure the successful execution of the PE samples in the dataset. The snapshot feature of the virtual machine is leveraged to roll it back after execution to ensure the uniformity of the software running environment. Besides, Cuckoo simulates some user actions (such as clicking a button, typing some text, etc.) to trigger the malicious behavior of malware. In this paper, we set the maximum running time of each sample to 5 minutes. That is to say, the sandbox process completes when the uploaded sample ends itself or runs for 5 minutes. After a PE file is uploaded, the Cuckoo server begins to call a free client to execute the file and record the API calls automatically. When the process completes, the Cuckoo server will generate a sandbox report about this uploaded file, and the raw API sequence can be extracted from this report.

\subsection{Evaluated Malware Detectors}
We employ some representative DNNs, i.e., LSTM \cite{staudemeyer2019understanding}, Text-CNN \cite{DBLP:conf/emnlp/Kim14}, and Transformer \cite{transformer} to build the malware detection models. These models learn the sequence features from API sequences and have been proven to be effective in malware detection. In fact, many existing studies have already used these three models or their variants or combinations as the encoder \cite{KolosnjajiZWE16, icassp, DBLP:journals/peerj-cs/CatakYEA20, DBLP:journals/compsec/LiLLWSQ22, DBLP:journals/tifs/ChenHLCZDL22}. The details of two DNN models are illustrated as follows:

\subsubsection{LSTM}
LSTM \cite{staudemeyer2019understanding} is a recurrent neural network architecture. It is able to capture the long-term context information through several gates designed to control the information transmission status. In this paper, we use the architecture of a single-layer LSTM in \cite{DBLP:journals/peerj-cs/CatakYEA20} as our baseline model. Specifically, we establish two LSTM models for comparison, namely the regular LSTM model and the LSTM model enhanced by MME. The regular model includes an embedding layer \cite{Embedding} to receive API name sequences as input, a single layer LSTM encoder, and an MLP classifier. The enhanced LSTM includes our API sequence embedding, a contrastive LSTM encoder, and a classifier with the same configuration as the regular model.
 
\begin{table}[]
	\centering
	\renewcommand\arraystretch{1}
	\caption{Comparisons of the regular and enhanced models (\%)}
	\label{Comparisons of LSTM and Text-CNN}
	\resizebox{\linewidth}{!}{
	\begin{tabular}{c|ccc|ccc|ccc}
		\toprule
		\multirow{2}{*}{\begin{tabular}[c]{@{}c@{}}Testing\\ Years\end{tabular}} & \multicolumn{3}{c|}{Regular LSTM}             & \multicolumn{3}{c|}{APIGraph(LSTM)\cite{DBLP:conf/ccs/ZhangZZDCZZY20}}    & \multicolumn{3}{c}{MME(LSTM)}                        \\
		\cmidrule(r){2-4} \cmidrule(r){5-7} \cmidrule(r){8-10}
		& FPR        & FNR  & F1   & FPR          & FNR  & F1  & FPR        & FNR           & F1           \\
		\cmidrule(r){1-1}\cmidrule(r){2-4} \cmidrule(r){5-7} \cmidrule(r){8-10}
		2018                                                                     & 6.52           & 21.19   & 84.75   & 6.91            & 15.81   & 87.80  & \textbf{5.96} & \textbf{8.04}    & \textbf{92.65}  \\
		2019                                                                     & \textbf{6.59} & 17.59   & 86.25   & 7.18            & 15.79   & 86.96  & 7.27          & \textbf{7.37}    & \textbf{91.61}  \\
		2020                                                                     & 10.38         & 24.22   & 83.16    & \textbf{9.55}   & 21.15   & 85.33  & 10.12         & \textbf{11.06}   & \textbf{91.10}  \\
		2021                                                                     & 8.96          & 26.57   & 81.78   & \textbf{8.15}   & 23.24   & 84.19  & 8.62          & \textbf{14.12}   & \textbf{89.55}  \\
		average                                                                  & 8.11        & 22.39 & 83.98 & \textbf{7.94} & 18.99  & 86.07  & 7.99        & \textbf{10.15} & \textbf{91.23} \\
		\cmidrule(r){1-1}\cmidrule(r){2-4} \cmidrule(r){5-7} \cmidrule(r){8-10}
		improve                                                                  & --             & --       & --       & \textcolor{green}{$\downarrow$0.17}  & 
        \textcolor{green}{$\downarrow$3.40} & \textcolor{red}{$\uparrow$2.09} & \textcolor{green}{$\downarrow$0.12}        & \textcolor{green}{$\downarrow$12.24} & 
        \textcolor{red}{$\uparrow$7.24}\\
		\bottomrule
		\toprule
		\multirow{2}{*}{\begin{tabular}[c]{@{}c@{}}Testing\\ Years\end{tabular}} & \multicolumn{3}{c|}{Regular Text-CNN} & \multicolumn{3}{c|}{APIGraph(Text-CNN)\cite{DBLP:conf/ccs/ZhangZZDCZZY20}} & \multicolumn{3}{c}{MME(Text-CNN)}                    \\
		\cmidrule(r){2-4} \cmidrule(r){5-7} \cmidrule(r){8-10}
		& FPR & FNR & F1   & FPR    & FNR     & F1      & FPR         & FNR          & F1           \\
		\cmidrule(r){1-1}\cmidrule(r){2-4} \cmidrule(r){5-7} \cmidrule(r){8-10}
		2018                                                                     & 5.19   & 19.45  & 86.50   & 5.62      & 12.90      & 90.13      & \textbf{2.39}  & \textbf{6.72}   & \textbf{95.23}  \\
		2019                                                                     & 5.47   & 21.27  & 84.70   & 5.73      & 12.11      & 89.93      & \textbf{4.38}  & \textbf{6.84}   & \textbf{93.65}  \\
		2020                                                                     & 5.87   & 24.97  & 83.98   & 7.18       & 17.01      & 88.54       & \textbf{5.39}  & \textbf{10.86}  & \textbf{92.62}  \\
		2021                                                                     & 6.38   & 25.12  & 83.54   & 6.81      & 19.68      & 86.83      & \textbf{4.17}  & \textbf{13.99}  & \textbf{91.08}  \\
		average                                                                  & 5.73   & 22.70  & 84.67 & 6.33     & 15.43    & 88.86    & \textbf{4.08}  & \textbf{9.60}   & \textbf{93.14}  \\
		\cmidrule(r){1-1}\cmidrule(r){2-4} \cmidrule(r){5-7} \cmidrule(r){8-10}
		improve                                                                  & --      & --      & --       & \textcolor{red}{$\uparrow$0.61}     & \textcolor{green}{$\downarrow$7.27}    & \textcolor{red}{$\uparrow$4.18}      & \textcolor{green}{$\downarrow$1.65} & 
        \textcolor{green}{$\downarrow$13.10} & \textcolor{red}{$\uparrow$8.47} \\
		\bottomrule
            \toprule
		\multirow{2}{*}{\begin{tabular}[c]{@{}c@{}}Testing\\ Years\end{tabular}} & \multicolumn{3}{c|}{Regular Transformer} & \multicolumn{3}{c|}{APIGraph(Transformer)\cite{DBLP:conf/ccs/ZhangZZDCZZY20}} & \multicolumn{3}{c}{MME(Transformer)}                    \\
		\cmidrule(r){2-4} \cmidrule(r){5-7} \cmidrule(r){8-10}
		& FPR & FNR & F1   & FPR    & FNR     & F1      & FPR         & FNR          & F1           \\
		\cmidrule(r){1-1}\cmidrule(r){2-4} \cmidrule(r){5-7} \cmidrule(r){8-10}
		2018                                                                     & \textbf{1.67}   & \textbf{6.56}  & 95.70   & 1.71      & 6.65      & \textbf{95.78}      & 1.82  & 6.66  & 95.57  \\
		2019                                                                     & 3.13   & 7.23  & 94.23   & 3.13      & 7.37      & \textbf{95.01}    & \textbf{2.63}  & \textbf{6.91}   & 94.72  \\
		2020                                                                     & 4.01   & 11.75  & 92.53   & 3.61       & \textbf{10.85}      & 91.46       & \textbf{3.36}  & 11.44  & \textbf{92.90}  \\
		2021                                                                     & 3.06   & 15.51  & 90.56   & 2.74      & 15.81      & 89.48      & \textbf{2.73}  & \textbf{14.81}  & \textbf{91.08}  \\
		average                                                                  & 2.97   & 10.26  & 93.25   & 2.80     & 10.17    & 92.93    & \textbf{2.63}  & \textbf{9.95}   & \textbf{93.57}  \\
		\cmidrule(r){1-1}\cmidrule(r){2-4} \cmidrule(r){5-7} \cmidrule(r){8-10}
		improve                                                                  & --      & --      & --       & \textcolor{green}{$\downarrow$0.17}     & \textcolor{green}{$\downarrow$0.09}    & \textcolor{green}{$\downarrow$0.32}      & \textcolor{green}{$\downarrow$0.1} & 
        \textcolor{green}{$\downarrow$0.31} & \textcolor{red}{$\uparrow$0.32} \\
		\bottomrule
	\end{tabular}
}
\vspace{-1.0em}
\end{table}
\subsubsection{Text-CNN}
Text-CNN \cite{DBLP:conf/emnlp/Kim14} is a variant of CNN used for text classification tasks. The regular model here also uses an embedding layer \cite{Embedding} and receives API name sequences as input. The filter size in CNN, or the n-gram size, denotes the number of successive API calls where the features are extracted. In the regular encoder, we set the filter sizes to 3, 4, and 5, respectively for three different Text-CNN layers. The enhanced Text-CNN includes our API sequence embedding, a contrastive Text-CNN encoder, and a classifier with the same configuration as the regular model.

\subsubsection{Transformer}
Transformer \cite{transformer} is a neural network architecture based on self-attention mechanisms, originally designed for natural language processing tasks. It excels at capturing long-range dependencies and parallelizing computations, making it highly effective for sequence modeling. In this paper, we use a Transformer model as a baseline for comparison. The regular Transformer model includes an embedding layer that receives API name sequences as input, a Transformer encoder with multiple self-attention layers, and an MLP classifier. The enhanced Transformer model incorporates our API sequence embedding, a contrastive Transformer encoder, and a classifier with the same configuration as the regular model.

\section{Evaluation}
In this section, we evaluate the effectiveness of MME in enhancing API sequence-based detection models. 


\subsection{Model Sustainability Analysis}
In this section, we measure the performance of existing malware detection models with and without the help of MME to understand the ability of MME in mitigating model degradation.
\subsubsection{Experimental Settings}
To evaluate the models' sustainability, we test the malware detectors yearly. For each detectors, we train a model on the samples of 2017, and sequentially test its performance on each year from 2018 to 2021. To ensure the effectiveness of the models, we employ a 10-fold cross-validation during the model training process and ensure that the all the models achieve an average F1 score of over 97\% on the validation set. During the model test, we calculate the false positive rate (FPR), false negative rate (FNR), and F1 score to evaluate how MME can help prolong the life-time of regular models. 

We also consider a state-of-the-art work called APIGraph \cite{DBLP:journals/tdsc/ZhangZZZZCY23},\cite{DBLP:conf/ccs/ZhangZZDCZZY20}, which is most relevant to our MME model, for comparison. APIGraph also leverages API knowledge graph learning and API clustering to enhance the regular malware detectors by capturing the semantically-equivalent APIs among evolved malware, thus slowing down the model aging. In fact, APIGraph primarily enhances the API name embedding stage of the model, whereas in comparison, our framework MME enhances both API name and argument embedding, as well as the encoder module.
 
\subsubsection{Results}
Table \ref{Comparisons of LSTM and Text-CNN} shows the performance of each baseline model in every test year. The phenomenon of model aging is observed quite prominently, especially in terms of the FNR. Over a four-year testing period, the regular LSTM model exhibited an average FNR as high as 22.39\%, resulting in a decrease in the F1 score to 83.98\%. Similarly, the regular Text-CNN model showed an average FNR of 22.70\%, with an accompanying drop in the F1 score to 84.67\%. This indicates a severe issue of elevated false negatives caused by the evolution of malicious software, as the model tends to classify unknown malware as benign.

Our enhancement method MME demonstrates significant improvement. Compared to the regular models, the MME(LSTM) exhibits a 12.24\% reduction in FNR and a 7.24\% increase in F1 score, with the average F1 value remaining above 90\%. The MME(Text-CNN), on the other hand, experiences a 13.10\% decrease in FNR and an 8.47\% increase in F1 score, and maintains an average F1 above 93\%. Moreover, in comparison to the state-of-the-art model APIGraph, our model lags behind by only 0.05\% in LSTM's FPR, while outperforming APIGraph in other metrics. These results indicate that our enhancement method possesses a strong ability to alleviate model aging.

The Transformer section of the table compares three models from 2018 to 2021. MME(Transformer) performs best, with the lowest average FPR (2.63\%) and FNR (9.95\%) and highest F1 (93.57\%), improving FPR by 0.34\%, FNR by 0.31\%, and F1 by 0.32\% over the regular Transformer (2.97\% FPR, 10.26\% FNR, 93.25\% F1). APIGraph(Transformer) shows slight improvements, with 2.80\% FPR, 10.17\% FNR, and 92.93\% F1, but is outperformed by MME(Transformer), which excels in reducing errors and boosting accuracy.

\subsection{Model Maintainability Analysis}
The purpose of this experiment is to evaluate how many human efforts MME can save while maintaining a high-performance malware detection model.

Specifically, the comparison includes two aspects. On the one hand, we compare the amount of human effort needed for active learning in maintaining both the regular and the enhanced models. On the other hand, we compare the model performance improvement given a fixed level of human effort.

\subsubsection{Comparison of human efforts needed to achieve a fixed performance}
First, we train a detection model on the samples of 2017, and test it month by month from Jan 2018 to Dec 2021. Then, when the F1 score of the model falls below a threshold $T$, we retrain the model so that it can reach the $T$. We calculate how many human efforts (i.e. the number of samples to label) are needed in the retraining step. To retrain an aged model, we adopt the active learning \cite{DBLP:conf/uss/PendleburyPJKC19} method, which is an optimization to normal retraining methods. Specifically, the uncertain sampling \cite{DBLP:conf/uss/PendleburyPJKC19} algorithm is used to actively select the most uncertain predictions. In detail, first we select the most 1\% uncertain samples to retrain the model, and then gradually increase the percentage by 1\% until the F1 score reaches $T$. Through this way, we can figure out the minimum efforts to maintain a high-performance model.

Table \ref{fixed retrain thresholds} shows the number of samples to label from 2018 to 2021 with $T = 0.95$ for both the regular and the enhanced models. It is clear that the models enahanced by MME can significantly save human efforts while reaching the threshold of $T$. For the LSTM model, the enhanced model can save 24.19\% to 70.27\% of human efforts during maintenance, with an average savings of 44.17\% over 4 years. Moreover, for the Text-CNN model, the enhanced model can save 33.53\% to 94.42\% of human efforts during maintenance, with an average savings of 52.12\% over 4 years. These results indicate that MME can significantly reduce human efforts when maintaining various malware detectors.

\begin{table}[]
	\centering
        \footnotesize
	\renewcommand\arraystretch{1}
	\caption{the number of labeled samples for active learning with fixed retrain thresholds ($F1 = 95\%$)}
	\label{fixed retrain thresholds}
	\begin{tabular}{ccccccc}
		\toprule
		\multirow{2}{*}{\begin{tabular}[c]{@{}c@{}}Testing\\ Years\end{tabular}} & \multicolumn{3}{c}{LSTM \# labeled samples} & \multicolumn{3}{c}{Text-CNN \# labeled samples} \\
		\cmidrule(r){2-4} \cmidrule(r){5-7}
		& Regular      & MME    & improve    & Regular  & MME  & improve  \\
		\cmidrule(r){1-1}\cmidrule(r){2-4} \cmidrule(r){5-7}
		2018                                                                   & 1,729     & 514           & 70.27\%    & 735       & 41              & 94.42\%  \\
		2019                                                                   & 1,645     & 1,247         & 24.19\%    & 1,265     & 662             & 47.68\%  \\
		2020                                                                   & 5,462     & 3,113         & 43.01\%    & 2,977     & 1,272           & 57.27\%  \\
		2021                                                                   & 3,402     & 1,959         & 42.41\%    & 2,195     & 1,459           & 33.53\%  \\
		\cmidrule(r){1-1}\cmidrule(r){2-4} \cmidrule(r){5-7}
		Total                                                                  & 12,238    & 6,833         & 44.17\%    & 7,172     & 3,434           & 52.12\%  \\
		\bottomrule
	\end{tabular}
        \vspace{-1.0em}
\end{table}

\subsubsection{Comparison of model performance improvement given a fixed level of human efforts}
The second comparison setting is to fix the amount of human effort and test the model performance of the regular and enhanced models. Similarly, we train a detector with samples from 2012, and test the detector month by month from Jan 2018 to Dec 2021. We also use the uncertain sampling \cite{DBLP:conf/uss/PendleburyPJKC19} in this experiment. We adopt two fixed human effort strategies: the first one is sample budgeting, where 20, 50, and 100 samples are labeled and used for retraining in each month; the second one is ratio budgeting, where 1\%, 5\%, and 10\% of the samples from each month are labeled and used for retraining. Finally, we calculate the FPR, FNR, and F1 score of the model in each month, and calculate their respective averages as the final comparison metrics.

As shown in Table \ref{fixed monthly sample labeling budget} and Table \ref{fixed monthly ratio labeling budget}, it can be observed that under the same level of human efforts, the enhanced models achieve better performance. Although there are instances where the FPR results may slightly increase compared to the regular models, this increase is less than 1\%. Significant improvements are seen in FNR and F1 scores, particularly in the reduction of FNR. Especially when fixing the analysts labeling effort at a low standard (such as 20 or 1\% samples per month), the models enhanced with MME show even more significant performance improvements compared to the regular models, where the FNR can be reduced by more than 10\%. This implies that the enhanced models, with just a slight amount of human efforts, can significantly mitigate the impact of malware evolution. The result also indicates that under a fixed level of human efforts, the models enhanced with MME achieve better performance, particularly in reducing FNR and improving F1 score.

For both LSTM and Text-CNN, using the models enhanced with MME, only 20 samples or 1\% of samples need to be labeled each month to keep the FNR below 10\% and achieve an F1 score above 90\%. In contrast, the regular models in our experiment require to label around 100 samples or 5\% of the samples per month to achieve the same effect. This experimental result indicates that MME can reduce the analysts labeling effort by 5 times. 

\begin{table}[]
	\centering
	\renewcommand\arraystretch{1}
	\caption{Active learning with a fixed monthly sample labeling budget}
	\label{fixed monthly sample labeling budget}
	\resizebox{\linewidth}{!}{
	\begin{tabular}{c|c|c|ccc}
		\toprule
		\multirow{2}{*}{\begin{tabular}[c]{@{}c@{}}Monthly \\ Sample Budget\end{tabular}} & \multirow{2}{*}{\begin{tabular}[c]{@{}c@{}}Base \\ Model\end{tabular}} & \multirow{2}{*}{Method} & \multicolumn{3}{c}{Average Performance}  \\
		\cmidrule(r){4-6}
		&                                                                          &                         & FPR(\%)      & FNR(\%)      & F1(\%)                 \\
		\midrule
		\multirow{6}{*}{20}                                                               & \multirow{3}{*}{LSTM}                                                    & Regular                 & 6.96   & 21.24  & 85.08            \\
		\cmidrule(r){3-3} \cmidrule(r){4-6}
		&                                                                          & \multirow{2}{*}{MME}                    & 7.73   & 9.19   & 91.86            \\
		&                                                                          &                         & \textcolor{red}{$\uparrow$0.77}  & \textcolor{green}{$\downarrow$12.05}  & \textcolor{red}{$\uparrow$6.77} \\
		\cmidrule(r){2-3} \cmidrule(r){4-6}
		& \multirow{3}{*}{Text-CNN}                                                & Regular                 & 5.50   & 20.87  & 85.91            \\
		\cmidrule(r){3-3} \cmidrule(r){4-6}
		&                                                                          & \multirow{2}{*}{MME}                    & 3.68   & 8.76   & 93.77            \\
		&                                                                          &                         & \textcolor{green}{$\downarrow$1.82} & \textcolor{green}{$\downarrow$12.10} & \textcolor{red}{$\uparrow$7.87} \\
		\midrule
		\multirow{6}{*}{50}                                                               & \multirow{3}{*}{LSTM}                                                    & Regular                 & 5.42   & 15.68  & 89.05            \\
		\cmidrule(r){3-3} \cmidrule(r){4-6}
		&                                                                          & \multirow{2}{*}{MME}                    & 6.40   & 6.79   & 93.71            \\
		&                                                                          &                         & \textcolor{red}{$\uparrow$0.98}  & \textcolor{green}{$\downarrow$8.88} & \textcolor{red}{$\uparrow$4.66} \\
		\cmidrule(r){2-3} \cmidrule(r){4-6}
		& \multirow{3}{*}{Text-CNN}                                                & Regular                 & 2.54   & 15.99  & 90.10            \\
		\cmidrule(r){3-3} \cmidrule(r){4-6}
		&                                                                          & \multirow{2}{*}{MME}                    & 3.28   & 7.93   & 94.40            \\
		&                                                                          &                         & \textcolor{red}{$\uparrow$0.74}  & \textcolor{green}{$\downarrow$8.06}  & \textcolor{red}{$\uparrow$4.30}      \\
		\midrule
		\multirow{6}{*}{100}                                                              & \multirow{3}{*}{LSTM}                                                    & Regular                 & 5.68   & 10.88  & 91.65            \\
		\cmidrule(r){3-3} \cmidrule(r){4-6}
		&                                                                          & \multirow{2}{*}{MME}                    & 4.80   & 5.83   & 94.90            \\
		&                                                                          &                         & \textcolor{green}{$\downarrow$0.88} & \textcolor{green}{$\downarrow$5.05} & \textcolor{red}{$\uparrow$3.25}   \\
		\cmidrule(r){2-3} \cmidrule(r){4-6}
		& \multirow{3}{*}{Text-CNN}                                                & Regular                 & 4.00   & 8.67   & 93.56            \\
		\cmidrule(r){3-3} \cmidrule(r){4-6}
		&                                                                          & \multirow{2}{*}{MME}                    & 3.28   & 6.73   & 95.04            \\
		&                                                                          &                         & \textcolor{green}{$\downarrow$0.72} & \textcolor{green}{$\downarrow$1.95} & \textcolor{red}{$\uparrow$1.48} \\
		\bottomrule
	\end{tabular}
}
\vspace{-1.0em}
\end{table}

\subsection{Model Ablation Analysis}
In this experiment, we want to measure the individual effects of the two parts of the MME framework (i.e., embedding enhancement and encoder enhancement) on enhancing the regular model.

\subsubsection{Experimental Settings}
\begin{figure*}[t]
	\centering
	\includegraphics[width=\textwidth]{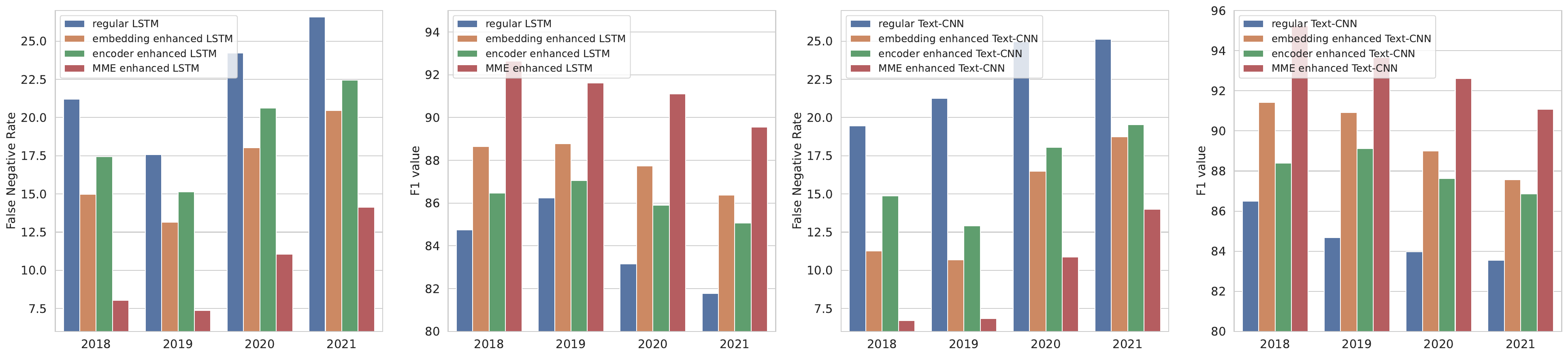}
	\caption{Model ablation analysis for LSTM \& TextCNN.}
	\label{ablation_study}
\end{figure*}

\begin{figure}[t]
\centering
\includegraphics[scale=0.277]{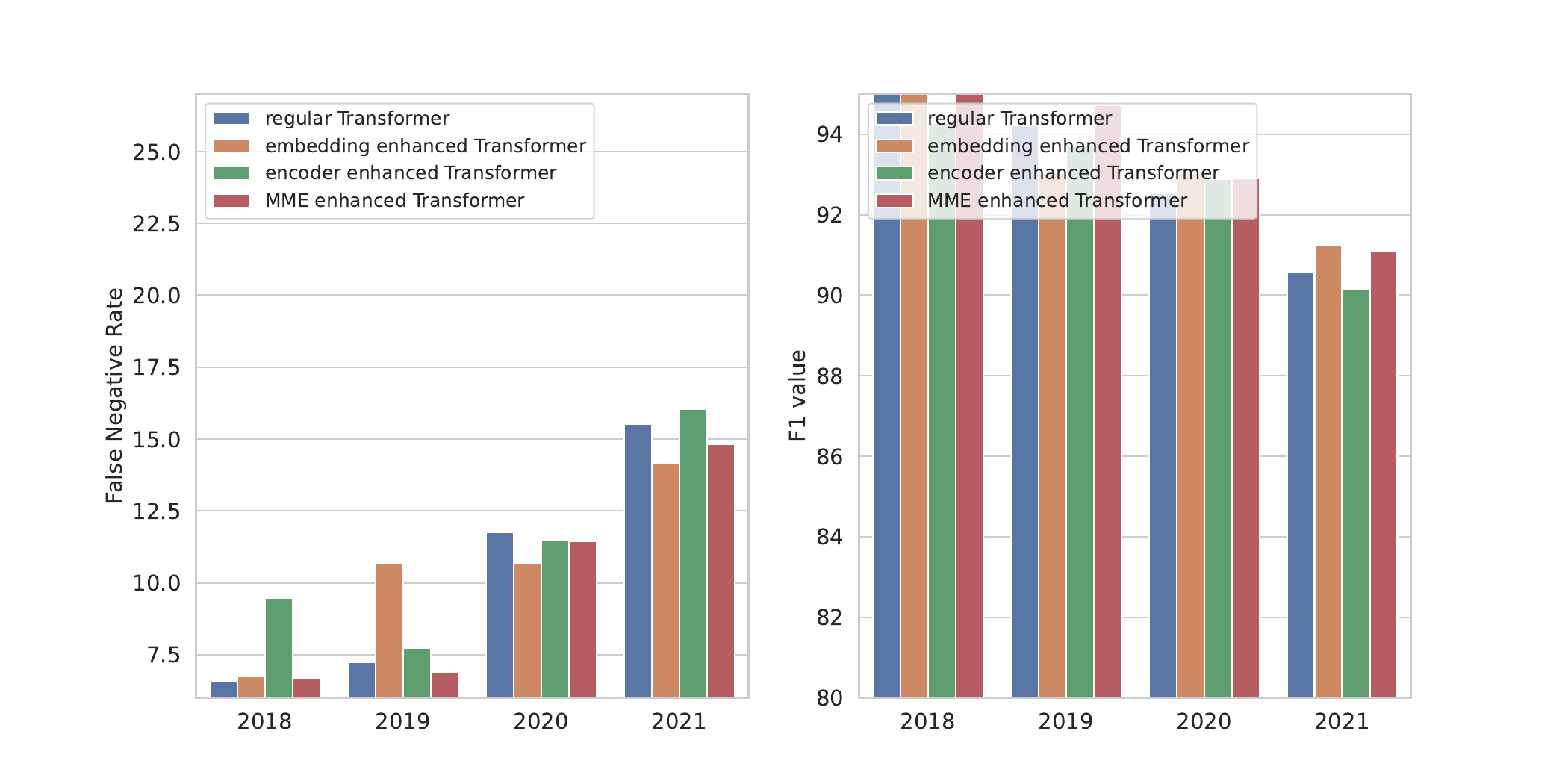}
\caption{Model ablation analysis for Transformer}
\label{Ablation_transformer}
\end{figure}

In the MME framework we proposed, there are two enhanced components: embedding enhancement and encoder enhancement. The embedding enhancement consists of API name embedding in \S III and API argument embedding in \S IV. The encoder enhancement refers to the contrastive encoder in \S V. To evaluate the impact of each component on the regular model's enhancement, we construct two MME variants: one with only embedding enhancement and another with only encoder enhancement. We train the 4 models (one regular model, two MME variants, and one proposed MME enhanced model) on the samples of 2017, and test their performance on each year from 2018 to 2021. Based on the previous experiments, it is evident that the main indicators of decreased model performance are the increase in false negative rates and the decrease in F1 scores. Therefore, we use these two metrics to assess the influence of each part of the MME framework on the model's enhancement. Figure \ref{ablation_study} and Figure \ref{Ablation_transformer} show results of the ablation experiments, where the baseline models consist of LSTM, Text-CNN, and Transformer with the same experimental settings as \S VI-C. The embedding enhanced LSTM, Text-CNN, and Transformer refer to the variant with only embedding enhancement, while the encoder enhanced LSTM/Text-CNN refers to the variant with only encoder enhancement. 

\begin{table}[]
	\centering
	\renewcommand\arraystretch{1}
	\caption{Active learning with a fixed monthly ratio labeling budget}
	\label{fixed monthly ratio labeling budget}
	\resizebox{\linewidth}{!}{
		\begin{tabular}{c|c|c|ccc}
			\toprule
			\multirow{2}{*}{\begin{tabular}[c]{@{}c@{}}Monthly \\ Ratio Budget\end{tabular}} & \multirow{2}{*}{\begin{tabular}[c]{@{}c@{}}Base \\ Model\end{tabular}} & \multirow{2}{*}{Method} & \multicolumn{3}{c}{Average Performance}  \\
			\cmidrule(r){4-6}
			&                                                                          &                         & FPR(\%)      & FNR(\%)      & F1(\%)                 \\
			\midrule
			\multirow{6}{*}{1\%}                                                               & \multirow{3}{*}{LSTM}                                                    & Regular                 & 6.48   & 19.51  & 86.32            \\
			\cmidrule(r){3-3} \cmidrule(r){4-6}
			&                                                                          & \multirow{2}{*}{MME}                    & 6.66   & 8.35   & 92.76            \\
			&                                                                          &                         & \textcolor{red}{$\uparrow$0.18}  & \textcolor{green}{$\downarrow$11.16}  & \textcolor{red}{$\uparrow$6.44}            \\
			\cmidrule(r){2-3} \cmidrule(r){4-6}
			& \multirow{3}{*}{Text-CNN}                                                & Regular                 & 5.33   & 19.10  & 87.05            \\
			\cmidrule(r){3-3} \cmidrule(r){4-6}
			&                                                                          & \multirow{2}{*}{MME}                    & 3.78   & 8.55   & 93.84            \\
			&                                                                          &                         & \textcolor{green}{$\downarrow$1.55} & \textcolor{green}{$\downarrow$10.55} & \textcolor{red}{$\uparrow$6.79} \\
			\midrule
			\multirow{6}{*}{5\%}                                                               & \multirow{3}{*}{LSTM}                                                    & Regular                 & 6.41   & 9.92  & 91.86            \\
			\cmidrule(r){3-3} \cmidrule(r){4-6}
			&                                                                          & \multirow{2}{*}{MME}                    & 4.28   & 6.06   & 95.00            \\
			&                                                                          &                         & \textcolor{green}{$\downarrow$2.13}  & \textcolor{green}{$\downarrow$3.86} & \textcolor{red}{$\uparrow$3.14} \\
			\cmidrule(r){2-3} \cmidrule(r){4-6}
			& \multirow{3}{*}{Text-CNN}                                                & Regular                 & 3.97   & 8.46  & 93.69            \\
			\cmidrule(r){3-3} \cmidrule(r){4-6}
			&                                                                          & \multirow{2}{*}{MME}                    & 3.64   & 6.03   & 95.25            \\
			&                                                                          &                         & \textcolor{green}{$\downarrow$0.34}  & \textcolor{green}{$\downarrow$2.43}  & \textcolor{red}{$\uparrow$1.56}            \\
			\midrule
			\multirow{6}{*}{10\%}                                                              & \multirow{3}{*}{LSTM}                                                    & Regular                 & 5.28   & 8.01  & 93.38            \\
			\cmidrule(r){3-3} \cmidrule(r){4-6}
			&                                                                          & \multirow{2}{*}{MME}                    & 3.90   & 5.45   & 95.49            \\
			&                                                                          &                         & \textcolor{green}{$\downarrow$1.39} & \textcolor{green}{$\downarrow$2.56} & \textcolor{red}{$\uparrow$2.11}            \\
			\cmidrule(r){2-3} \cmidrule(r){4-6}
			& \multirow{3}{*}{Text-CNN}                                                & Regular                 & 3.31   & 4.98   & 95.63            \\
			\cmidrule(r){3-3} \cmidrule(r){4-6}
			&                                                                          & \multirow{2}{*}{MME}                    & 2.35   & 4.33   & 96.70            \\
			&                                                                          &                         & \textcolor{green}{$\downarrow$0.96} & \textcolor{green}{$\downarrow$0.65} & \textcolor{red}{$\uparrow$1.07} \\
			\bottomrule
		\end{tabular}
	}
\vspace{-1.0em}
\end{table}

\subsubsection{Results}
Intuitively, both embedding and encoder enhancements have demonstrated significant improvements to the model, suggesting that optimizing API sequence embeddings through contrastive learning and refining the encoder’s training process can effectively mitigate the impact of malware evolution. Further observations reveal differences in the effects of these enhancements. Over the four-year testing period from 2018 to 2021, for the LSTM model, the embedding-enhanced LSTM model exhibited an average decrease of 5.7\% in FNR and an average increase of 3.9\% in F1 score compared to the regular LSTM model. In contrast, the encoder-enhanced LSTM model showed an average decrease of 3.5\% in FNR and an average increase of 2.1\% in F1 score. For the Text-CNN model, the embedding-enhanced Text-CNN achieved an average decrease of 8.4\% in FNR and an average increase of 5\% in F1 score compared to the regular Text-CNN, while the improved Text-CNN model displayed an average decrease of 6.4\% in FNR and an average increase of 3.3\% in F1 score. For the Transformer model, compared to the regular Transformer, our proposed MME framework model recorded an average decrease of 0.7\% in FPR and an average increase of 0.32\% in F1 score. Although this improvement is not substantial, it still reflects an overall enhancement. From these results, it is evident that embedding-enhanced outperforms encoder enhancement slightly, indicating that well-designed feature representations play a critical role in mitigating model aging. Finally, the proposed MME framework, which integrates both embedding and encoder enhancements, achieved the most effective mitigation. Over the four-year testing period, compared to the regular LSTM, the MME-enhanced LSTM demonstrated an average decrease of 12.24\% in FNR and an average increase of 7.2\% in F1 score. Similarly, compared to the regular Text-CNN, the MME-enhanced Text-CNN showed an average decrease of 13.10\% in FNR and an average increase of 8.47\% in F1 score. Notably, the embedding-enhanced Transformer exhibited the most significant reduction in FNR.

\subsection{Malware Feature Stability Analysis}

We observed that the malware evolution can disturb the stability of the original feature space, leading to a decline in model performance. In this experiment, we want to measure the stability of the feature space concerning the evolution of malware from the same family to show that the MME-enhanced model can capture the semantic similarity between the original and evolved of malware. 

\subsubsection{Experimental Settings}
Here is our evaluation methodology, which involves four steps. First, we select the top 10 malware families with the most number of samples from the dataset in \S VI-A. As a result, we have 17,288 malware samples in this experiment and every family has more than 1k samples. Second, for each malware family, we sort all the family samples by their appearing time and then divide them into 10 groups so that each group contains 10\% of the family. The samples in one group is strictly ahead of the samples from the next group in terms of their appearance time. Third, for each malware sample, we input its raw API sequence into the regular/MME-enhanced model and take the output of the regular/contrastive encoder as its feature representation. Lastly, we calculate a feature stability score of every two adjacent groups using Jensen–Shannon divergence \cite{menendez1997jensen}. The Jensen-Shannon divergence is a method used to measure the similarity between two feature distributions: $J S\left(P_1 \| P_2\right)=\frac{1}{2} K L\left(P_1 \| \frac{P_1+P_2}{2}\right)+\frac{1}{2} K L\left(P_2 \| \frac{P_1+P_2}{2}\right)$. It calculates the average Kullback-Leibler divergence between the two distributions (i.e., $KL(P_1 \| P_1) = \sum_{i} P_1(i) \log\left(\frac{P_1(i)}{P_2(i)}\right)$) and derives a final measurement value by utilizing the symmetry of the logarithmic function. In this experiment, $P_1$ and $P_2$ refer to the set of softmax-normalized features obtained in the third step for two adjacent groups. The score of $J S\left(P_1 \| P_2\right)$ ranges from 0 to 1, where the value closer to 0 indicates that the malware feature distributions between two groups are more similar, implying better feature space stability.

\subsubsection{Result}
\begin{figure}[t]
	\centering
	\includegraphics[width=\linewidth]{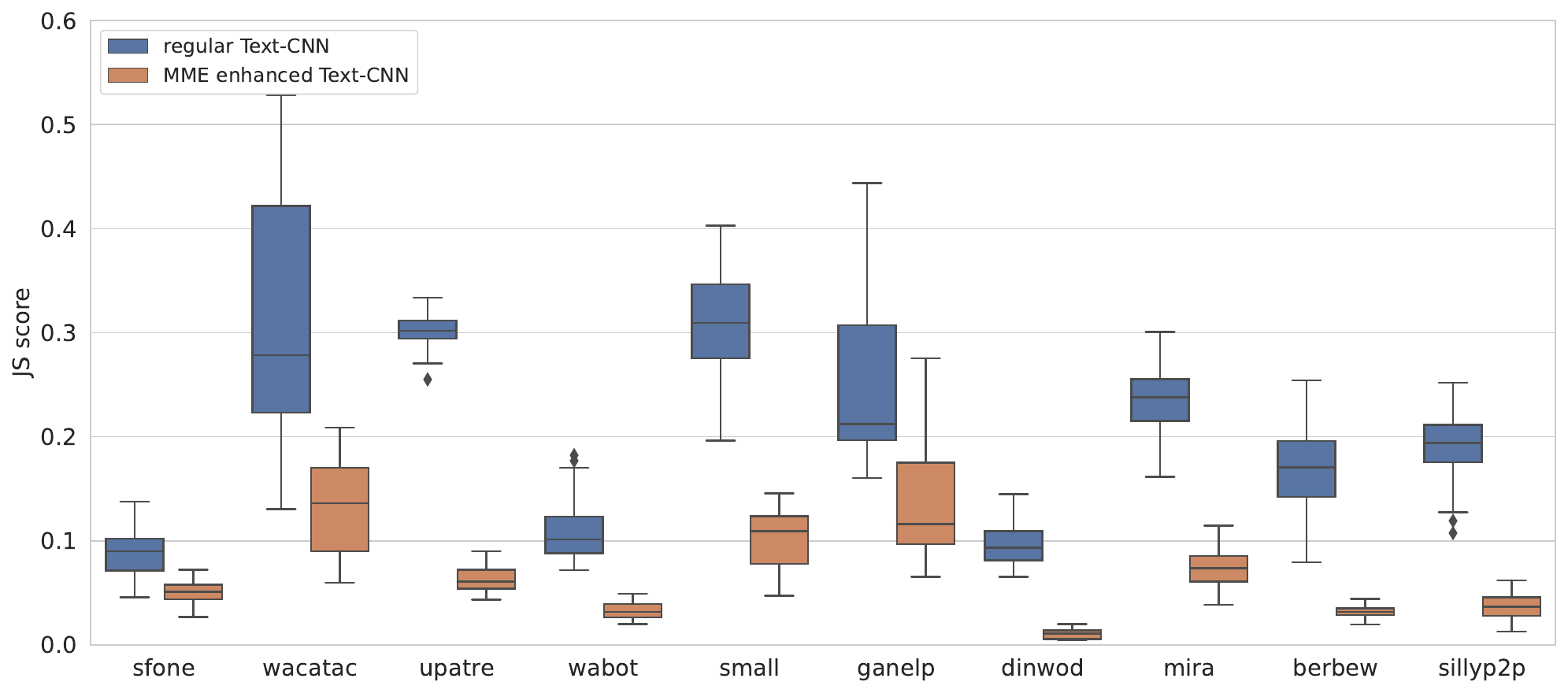}
	\caption{Feature stability analysis of Text-CNN model.}
	\label{JS_TextCNN}
        \vspace{-1.0em}
\end{figure}

\begin{figure}[t]
	\centering
	\includegraphics[width=\linewidth]{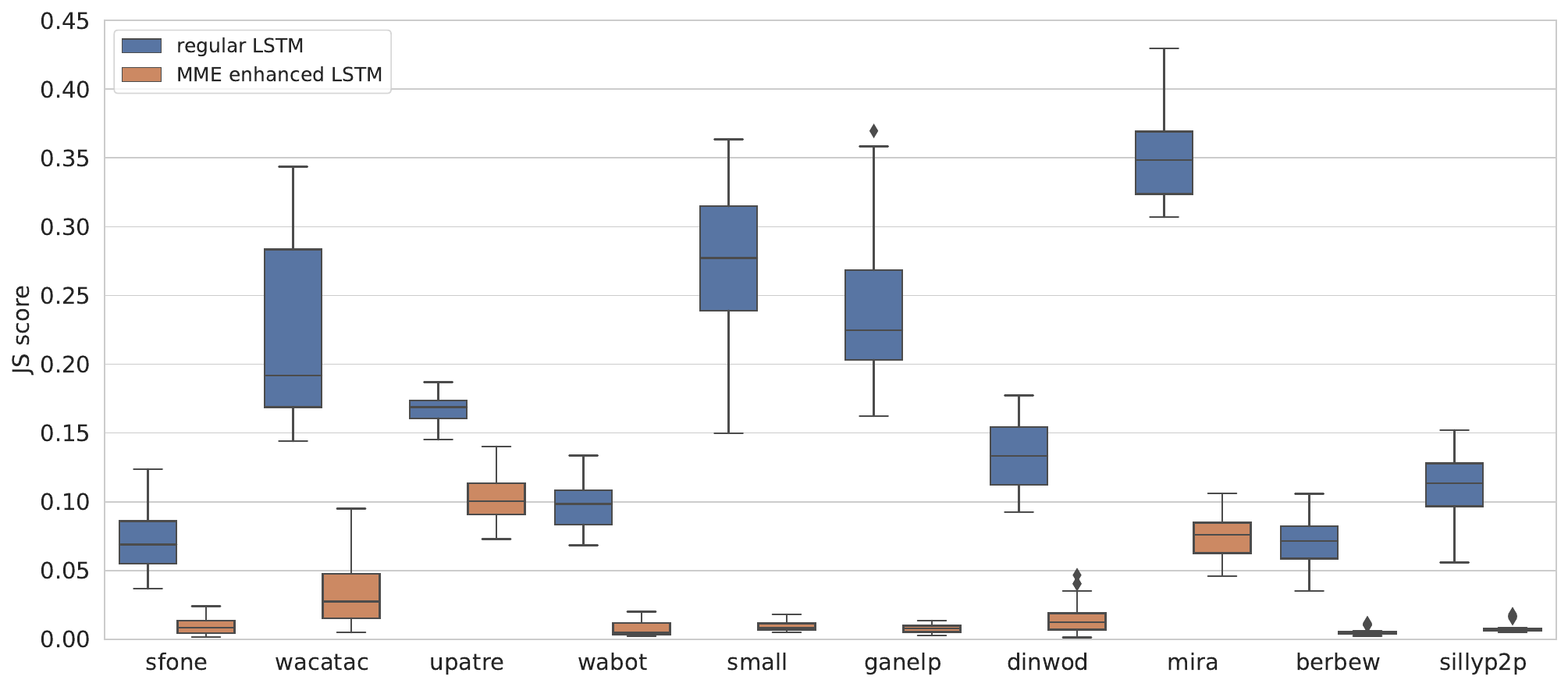}
	\caption{Feature stability analysis of LSTM model.}
	\label{JS_LSTM}
        \vspace{-1.0em}
\end{figure}

\begin{figure}[t]
	\centering
	\includegraphics[width=\linewidth]{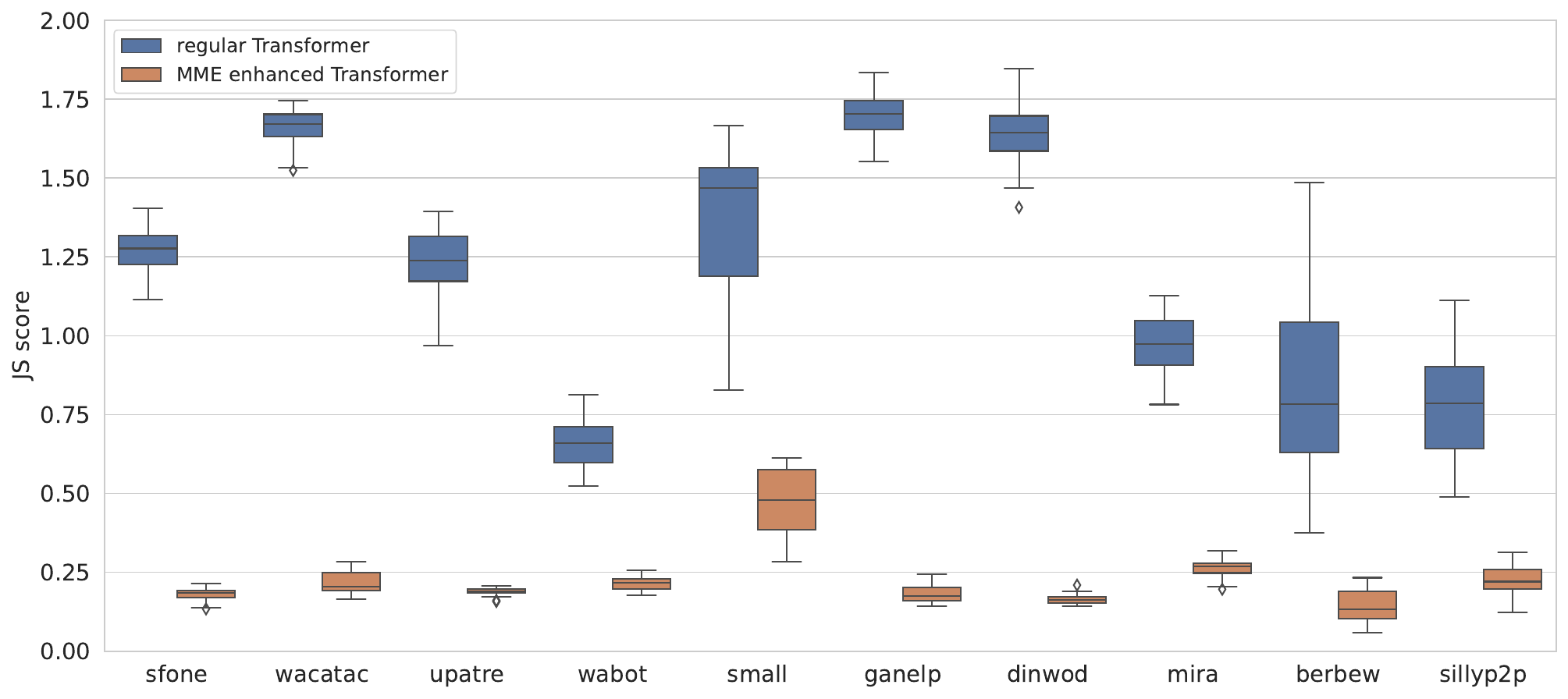}
	\caption{Feature stability analysis of Transformer model.}
	\label{fig:JS_Transformer}
        \vspace{-1.0em}
\end{figure}

Figure \ref{JS_TextCNN}, \ref{JS_LSTM} and \ref{fig:JS_Transformer} show the distribution of feature stability scores (i.e., $JS$ scores) for each malware family with the regular and MME-enhanced models. We can observe that for each malware family, the JS scores of all MME-enhanced models are closer to 0, significantly lower than the results of the regular model. This indicates that the feature stability of the MME-enhanced models demonstrates better performance. During the evolution of malware, the MME enhanced model can reduce the feature space disturbances. This experiment explains why MME can help the model mitigate the impact of malware evolution, as malware tends to retain semantic similarities during its evolution, and MME can capture these similarities and maintain the feature space stability.

\subsection{Comparison With Other Methods}

\begin{table*}
\caption{Performance Comparison of the Proposed Models with Other Methods (\%)}
\label{tb:MMEvsOtherModel}
\centering
\renewcommand{\arraystretch}{1.3}
\setlength{\tabcolsep}{3.6pt}
\begin{tabular}{c|c|cccc|cccc|cccc|cccc} 
\toprule
\multicolumn{2}{c|}{Dataset}                      & \multicolumn{4}{c|}{2018}                                                                                      & \multicolumn{4}{c|}{2019}                                                                                      & \multicolumn{4}{c|}{2020}                                                                                      & \multicolumn{4}{c}{2021}                                                                                       \\ 
\hline
\multicolumn{2}{c|}{Method}                       & Acc                       & F1                        & FPR                       & FNR                        & Acc                       & F1                        & FPR                       & FNR                        & Acc                       & F1                        & FPR                       & FNR                        & Acc                       & F1                        & FPR                       & FNR                        \\ 
\midrule
\multirow{5}{*}{MalScan}       & Degree           & 76.32                     & 66.67                     & 3.37                      & 47.98                      & 73.52                     & 53.44                     & 3.43                      & 61.64                      & 44.03                     & 61.14                     & 0.00                      & 55.97                      & 49.08                     & 65.84                     & 0.00                      & 50.92                      \\
                               & Closeness        & 77.18                     & 68.68                     & 4.24                      & 45.05                      & 74.20                     & 57.22                     & 5.70                      & 56.44                      & 46.82                     & 63.78                     & 0.00                      & 53.18                      & 51.46                     & 67.95                     & 0.00                      & 48.54                      \\
                               & Harmonic         & 76.09                     & 66.98                     & 4.85                      & 46.73                      & 72.67                     & 53.24                     & 5.43                      & 60.72                      & 44.82                     & 61.90                     & 0.00                      & 55.18                      & 51.06                     & 67.60                     & 0.00                      & 48.94                      \\
                               & Katz             & 75.40                     & 66.28                     & 5.98                      & 46.89                      & 72.40                     & 53.42                     & 6.31                      & 60.05                      & 45.60                     & 62.64                     & 0.00                      & 54.40                      & 51.06                     & 67.60                     & 0.00                      & 48.94                      \\
                               & Concatenate      & 76.60                     & 67.99                     & 4.99                      & 45.42                      & 73.81                     & 56.07                     & 5.46                      & 57.80                      & 46.78                     & 63.74                     & 0.00                      & 53.22                      & 53.46                     & 69.67                     & 0.00                      & 46.54                      \\ 
\hline
MalBERT                        & -                & 90.60                     & 89.86                     & 6.61                      & 12.48                      & 91.22                     & 89.67                     & 6.72                      & 11.51                      & 84.92                     & 87.04                     & 10.80                     & 17.75                      & 83.44                     & 85.04                     & 10.70                     & 20.58                      \\ 
\hline
Transformer                    & -                & 95.88                     & 95.70                     & 1.67                      & 6.56                       & 95.11                     & 94.23                     & 3.13                      & 7.23                       & 91.22                     & 92.53                     & 4.01                      & 11.75                      & 89.56                     & 90.56                     & 3.06                      & 15.51                      \\ 
\hline
API2Vec++                      & BiLSTM Attention            & \multicolumn{1}{l}{81.70} & \multicolumn{1}{l}{82.04} & \multicolumn{1}{l}{20.20} & \multicolumn{1}{l|}{16.40} & \multicolumn{1}{l}{87.45} & \multicolumn{1}{l}{87.41} & \multicolumn{1}{l}{11.90} & \multicolumn{1}{l|}{13.20} & \multicolumn{1}{l}{81.70} & \multicolumn{1}{l}{82.45} & \multicolumn{1}{l}{22.60} & \multicolumn{1}{l|}{14.00} & \multicolumn{1}{l}{83.90} & \multicolumn{1}{l}{84.32} & \multicolumn{1}{l}{18.80} & \multicolumn{1}{l}{13.40}  \\ 
\hline
DMalNet                        & GNNs             & 91.13                     & 90.32                     & 4.05                      & 13.25                      & 88.15                     & 89.32                     & 6.48                      & 15.92                      & 88.83                     & 87.39                     & 6.23                      & 19.08                      & 90.98                     & 90.01                     & 4.93                      & 14.96                      \\ 
\hline
\multirow{5}{*}{\textbf{Ours}} & MME(TextCNN)     & 95.34                     & 95.23                     & 2.39                      & 6.72                       & 94.22                     & 93.65                     & 4.38                      & \textbf{6.84}              & 92.51                     & 92.62                     & 5.39                      & 10.86                      & 91.83                     & 91.08                     & 4.17                      & 13.99                      \\
                               & MME(LSTM)        & 92.95                     & 92.65                     & 5.96                      & 8.04                       & 92.67                     & 91.61                     & 7.27                      & 7.37                       & 89.52                     & 91.10                     & 10.12                     & 11.06                      & 89.14                     & 89.55                     & 8.62                      & 14.12                      \\
                               & MME(DMalNet)     & 93.39                     & 92.71                     & 4.18                      & 8.81                       & 92.16                     & 93.63                     & 5.47                      & 9.64                       & \textbf{92.69}            & 92.63                     & 5.26                      & \textbf{10.6}              & \textbf{93.89}            & \textbf{93.60}            & 4.37                      & \textbf{8.64}              \\
                               & MME(Transformer) & \textbf{96.00}            & \textbf{95.57}            & \textbf{1.82}             & \textbf{6.66}              & \textbf{95.53}            & \textbf{94.72}            & \textbf{2.63}             & 6.91                       & 91.67                     & \textbf{92.90}            & \textbf{3.36}             & 11.44                      & 90.11                     & 91.08                     & \textbf{2.73}             & 14.81                      \\
                               & MME(BERT)        & 95.10                     & 94.74                     & 2.68                      & 7.34                       & 94.59                     & 93.65                     & 3.92                      & 7.38                       & 90.99                     & 92.36                     & 4.81                      & 11.63                      & 89.97                     & 91.02                     & 3.91                      & 14.24                      \\
\bottomrule
\end{tabular}
\end{table*}



DMalNet\cite{li2022dmalnet} exports an API call graph from a sequence of API calls, transforming the relationships between API calls into structural information of the graph. Finally, a graph neural network was used to realize malware detection and type classification. MalBERT\cite{xu2021malbert} proposes a static analysis of malicious code based on BERT\cite{bert}, which uses the preprocessed features to statically analyze the source code of malware, and divides the existing malicious code into different representative malicious code categories. API2Vec++ \cite{API2Vec++} designs a Temporal Process Graph (TPG) to model inter-process behavior and a Temporal API Property Graph (TAPG) to model intra-process behavior, and develops a heuristic random walk algorithm to generate numerous paths that can capture fine-grained malicious family behaviors. MalScan\cite{MalScan} targets Android malware, while MME focuses on Windows malware. However, we believe that a comparison in terms of methodology is valuable, so we also take MalScan\cite{MalScan} as a baseline. 

The experimental results, as shown in Table \ref{tb:MMEvsOtherModel}, demonstrate that our proposed MME model consistently and significantly outperforms other models across all four datasets from 2018 to 2021. Specifically, the MME(Transformer) model achieved the highest accuracy and F1 scores in 2018 and 2019, with an accuracy of 96.00\% and an F1 score of 95.57\% in 2018, and an accuracy of 95.53\% and an F1 score of 94.72\% in 2019. Furthermore, our MME(DMalNet) model exhibited outstanding performance on more recent datasets, achieving accuracies of 92.69\% in 2020 and 93.89\% in 2021. The effect of MME enhancing DMalNet verifies the ability of the MME framework to enhance GNN-based methods, and reflects the high adaptability and reliability of the MME framework. Compared to traditional feature extraction methods such as MalScan \cite{MalScan} and existing models like MalBERT \cite{xu2021malbert} and DMalNet \cite{li2022dmalnet}, our approach demonstrates clear improvements in both accuracy and F1 score. These results indicate that the proposed MME framework for malware analysis effectively enhances classification performance, reduces misclassification errors, and offers superior practical value.

Further analysis of misclassification rates among different models reveals that our proposed MME model significantly reduces both the FNR and FPR. For instance, on the 2018 dataset, our MME (Transformer) achieved a false positive rate of only 1.82\% and a false negative rate of 6.66\%, substantially lower than those of MalBERT \cite{xu2021malbert} (FPR of 6.61\%, FNR of 12.48\%) and DMalNet \cite{li2022dmalnet} (FPR of 4.05\%, FNR of 13.25\%). This advantage persisted consistently over subsequent years' datasets, demonstrating the superior stability and generalization capability of the MME model. This experiment comprehensively validates the advantages and effectiveness of our proposed MME architecture in terms of accuracy, reliability, and robustness, positioning it as a valuable method for current malware detection tasks.

\subsection{Knowledge Graph Analysis}
\begin{figure}[t]
    \centering
    \includegraphics[width=\linewidth]{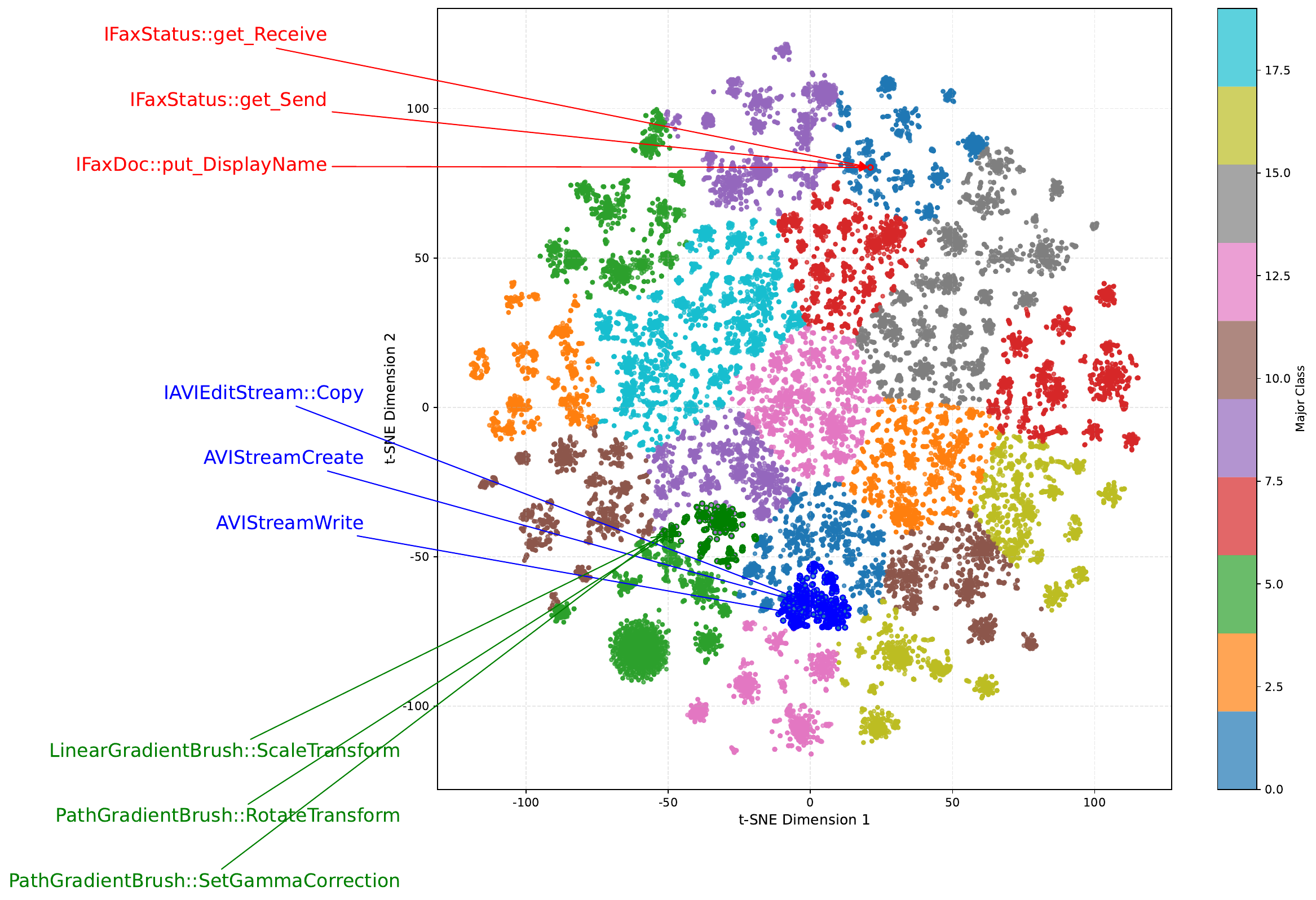}
    \caption{T-SNE Visualization API Entity Embedding}
    \label{fig:APIEntityVis}
    \vspace{-1.0em}
\end{figure}

We employed T-SNE for dimensionality reduction and K-means for clustering to visualize the API entity embeddings. The reduced data was subsequently grouped to identify the primary functional categories of APIs, resulting in 20 distinct categories. From each cluster, three API entities proximate to the centroid were randomly selected as representatives. The outcomes, illustrated in Figure \ref{fig:APIEntityVis}, highlight methods such as \textit{IFaxStatus::get\_Receive}, \textit{IFaxStatus::get\_Send}, and \textit{IFaxDoc::put\_DisplayName}, all of which pertain to fax-related operations. Observations indicate that the knowledge graph forms compact clusters, thereby facilitating the differentiation between benign and malicious software. This suggests that the knowledge graph effectively captures salient features of malware, establishing discernible classification boundaries within the feature space.

We conducted a centrality analysis on the knowledge graph and listed the different centrality measures along with their corresponding values, as presented in Table \ref{tb:centrality_analysis}. We identified the top two key API entities for each centrality type, leading to the following conclusions: \textit{ldap\_parse\_page\_control} and \textit{get\_RealTimeBuffersLost}, as well as \textit{SHCreateShellPalette}, serve as both central nodes and key bridges within the network. These entities function as critical hubs in the knowledge graph, characterized by numerous direct connections and high influence, which underscores their core status within the system. The knowledge graph elucidates their semantic and structural relationships within the network, thereby aiding in the differentiation between benign and malicious software.
\begin{table}[h]
\centering
\small
\caption{Results of the centrality analysis}
\label{tb:centrality_analysis}
\begin{tabular}{c|llll} 
\toprule
            & \multicolumn{4}{c}{API Name}                                                                                                           \\ 
\hline
Degree      & \multicolumn{4}{c}{\begin{tabular}[c]{@{}c@{}}ldap\_parse\_page\_control(0.0023)\\get\_RealTimeBuffersLost(0.0019)\end{tabular}}       \\ 
\hline
Closeness   & \multicolumn{4}{c}{\begin{tabular}[c]{@{}c@{}}SHCreateShellPalette(0.1230)\\~ICertPropertyArchived::Initialize(0.1224);\end{tabular}}  \\ 
\hline
Betweenness & \multicolumn{4}{c}{\begin{tabular}[c]{@{}c@{}}SHCreateShellPalette(0.0122)\\~IMFMediaSession::GetClock(0.0082);\end{tabular}}          \\ 
\hline
Eigenvector & \multicolumn{4}{c}{\begin{tabular}[c]{@{}c@{}}ldap\_parse\_page\_control(0.3060)\\get\_RealTimeBuffersLost(0.2271);\end{tabular}}      \\
\bottomrule
\end{tabular}
\end{table}

\section{Related Work}
\subsection{API Sequence-based Malware Detection} 




Dynamic malware detection executes the software in a secured virtual environment and monitors its run-time behavior. A running software calls many system APIs, which characterize software behaviors including network access, file creation and modification, etc. These API calls form an API call sequence which has become a widely used data source for malware detection and classification \cite{ChangZK22, DBLP:journals/compsec/UcciAB19, KolosnjajiZWE16, icassp, DBLP:journals/compsec/LiLLWSQ22, DBLP:journals/tifs/ChenHLCZDL22, DBLP:journals/peerj-cs/CatakYEA20,  DBLP:conf/acsac/RabadiT20, 2020Amer, zhang2020dynamic, tran2017nlp, kim2018ntmaldetect, li2022dmalnet}.

Inspired by deep learning-based sequence analysis, many researchers apply some DL models like convolutional neural networks (CNNs) and recurrent neural networks (RNNs) to learn features of the API call sequences. Kolosnjaji et al. \cite{KolosnjajiZWE16} use the API sequence as input. Their approach stacks a CNN that uses a 3-sized filter to represent 3 consecutive APIs (like the 3-gram approach). After the CNN, the LSTM is applied to handle the time-series sequence. Agrawal et al. \cite{icassp} input the API names and the n-gram of the string arguments into several stacked LSTMs. Zhang et al. \cite{zhang2020dynamic}\cite{ASDroid},\cite{MultimodalFF} build a feature engineering about the API names and arguments and then design a deep learning model including gate-CNNs and Bi-LSTM as the malware detector. Catak et al. \cite{DBLP:journals/peerj-cs/CatakYEA20} input API sequences into LSTMs to detect and classify malware. Li et al. \cite{DBLP:journals/compsec/LiLLWSQ22} combine the Text-CNN with Bi-LSTM to analyze the API sequences and detect malware. Similarly, Chen et al. \cite{DBLP:journals/tifs/ChenHLCZDL22} use the Text-CNN and Bi-LSTM as the baseline models for API sequence-based malware detection. 

Obviously, too many works have already using Text-CNN, LSTM and Transformer models or their variants or combinations as the encoder of the detection models. However, how to perform feature representation so that the model can accurately understand the semantic information in API sequences and thus understand the software behavior remains a challenging issue. Our framework MME enhance the API sequence embedding and can better represent each API as a semantic feature vector, as shown in \S \ref{sec:API Name Embedding} and \S \ref{sec:API Argument Embedding}.

\subsection{Model Aging Caused by Malware Evolution}


Deep learning techniques were originally designed for stationary environments in which the training and test sets are assumed to be generated from the same statistical distribution. However, this assumption is not valid in the malware domain. Malware samples, including various families, evolve over time due to changes resulting from adding capabilities, fixing bugs, porting to new platforms, etc. Thus, malware detectors are deployed in dynamic environments, where malware variants keep evolving, causing the performance to deteriorate significantly over time. This is known as the problem of model aging or concept drift \cite{DBLP:conf/uss/PendleburyPJKC19}.

There are mainly two methods to address model aging caused by the evolution of malware. The first is to retrain and update detection models with newly labeled samples, or reject drift samples until they can be expertly analyzed. For example, in Android malware detection, DroidEvolver \cite{DBLP:conf/eurosp/XuLDCX19} utilizes online learning and pseudo-labels to self-update the detection model. However, the accumulation of pseudo-label errors may lead to model self-poisoning which have catastrophic effects on performance. Some studies focus on detecting drift samples that deviate from existing classes from a large number of test samples and update models using periodical retraining \cite{DBLP:conf/uss/Yang0HCAX021, DBLP:conf/uss/JordaneySDWPNC17}. However, labeling samples and retraining the model still requires a lot of expert knowledge and computing resources. More importantly, it is also difficult to determine when the model should be retrained. Delayed retraining can leave the outdated model vulnerable to evolved malware.

The second method is to deliberately consider the issue of model aging during the process of model design and feature space optimization. Researchers represent features to be more robust against temporal bias and reduce the impact of malware evolution. APIGraph \cite{DBLP:conf/ccs/ZhangZZDCZZY20} is proposed to enhance state-of-the-art malware detectors with the similarity information among evolved malware in terms of equivalent or similar API usages. It constructs an API knowledge graph based on the API documentation, and use graph embedding and the K-means algorithm to cluster APIs with similar semantics. Similarity, SDAC \cite{DBLP:journals/tdsc/XuLDX22} calculates APIs’ contributions to malware detection and assign APIs to feature vectors. Then, SDAC clusters all APIs based on their semantic distances to create a feature set in the training phase, and extends the feature set to include all new APIs in the detecting phase. However, the detectors above mainly focus on API occurrence or API frequency, which is difficult to be applied to dynamic malware detection based on API sequence analysis. The MME proposed in this paper further extends API sequence embedding and constructs a contrastive encoder to address the model aging issue in API sequence-based malware detection.

\section{Discussion And Limitation}
\subsection{Other types of Detectors}
The API sequence is a popular type of feature widely adopted by dynamic malware detectors \cite{ChangZK22, DBLP:journals/compsec/UcciAB19, KolosnjajiZWE16, icassp, DBLP:journals/compsec/LiLLWSQ22, DBLP:journals/tifs/ChenHLCZDL22, DBLP:journals/peerj-cs/CatakYEA20,  DBLP:conf/acsac/RabadiT20, 2020Amer, zhang2020dynamic}, mainly because API sequences are essential in understanding malware behaviors. In our experiments, we validate the effectiveness of MME by enhancing the LSTM, Text-CNN, and other models. There indeed are many other DNN models, such as Bi-LSTM, Gate-CNN, or even more variant models commonly used for learning API sequences \cite{icassp, DBLP:journals/compsec/LiLLWSQ22, DBLP:journals/tifs/ChenHLCZDL22, zhang2020dynamic}. Although these models are not individually validated in this paper, they also require the API sequence embedding and encoder training process. We believe that the MME approach can also be applied to these models and achieve similar enhancement effects. Current coarse-grained methods for contrastive learning have certain limitations, and future work can propose work that explores hierarchical constraints and adversarial enhancement robustness.

\subsection{Overly Advanced Malware}
In reality, certain malware employs highly advanced attack methods, rendering their behavior nearly indistinguishable from previously known malware. For such emerging malware, the performance of MME may decline. In this scenario, drift sample detection methods\cite{DBLP:conf/uss/Yang0HCAX021, DBLP:conf/uss/JordaneySDWPNC17} can be utilized to identify these samples, as they significantly deviate from the original data distribution. Subsequently, these highly advanced malware samples can be collected and labeled to update the model. It is believed that integrating the MME enhancement model with periodic drift sample detection can more effectively address the ongoing evolution of malware. Furthermore, advanced evasion techniques, such as polymorphism, virtualization detection, and adversarial triggers, fall outside the experimental scope of this study. Regarding the knowledge graph, we recognize that extracting only the first verb from API description sentences may overlook other actions, potentially impacting functional similarity judgments. Future work could incorporate large language model techniques or more sophisticated NLP methods to refine this aspect. While MME’s knowledge graph and contrastive learning may offer some robustness against these threats, their effectiveness remains untested. Future research will explore MME’s performance in these scenarios by integrating adversarial training and drift detection mechanisms.

\subsection{Malware Sandbox Evasion}
To obtain the raw API sequences, sandboxes are widely used to execute malware inside virtual machines and monitor APIs with API hooking techniques. Sandbox evasion refers to techniques employed by malware to avoid detection or analysis within a sandbox environment \cite{DBLP:conf/raid/LindorferKC11}. For example, when malware detects that it is running within a sandbox environment, it can refrain from executing any malicious operations, or even disguise itself as a legitimate application to exhibit benign behavior. The key to combating such ``environment-aware" malware is to optimize the sandbox environment to closely resemble a real system environment. In the sandbox used for experiments, we simulate some user actions (such as clicking a button, typing some text, etc.) to trigger malware's real behaviors. We further utilize the statistical model proposed by Miramirkhani et al. \cite{miramirkhani2017spotless} to optimize and fine-tune the sandbox, making it even closer to a real system environment. We believe these operations can help the sandbox capture the real API sequences of malware. In future works, solutions \cite{galloro2022systematical, liu2022enhancing, d2020dissection, avllazagaj2021malware} focusing on detecting sandbox evasion can be used to further optimize this issue.

\section{Conclusion}
This paper proposes a model enhancement method, MME to mitigate the impact of malware evolution on API sequence-based windows malware detectors. We observe that the API sequences of malware samples before and after evolution usually have similar malicious semantics, including equivalent API usage, similar system resources, and similar API fragments. This provides an opportunity to reduce the feature gaps caused by evolution, and slowing down model aging. Firstly, by establishing an API knowledge graph and capture semantic similarities between APIs, the influence of equivalent API substitution is reduced. Secondly, by adopting a hierarchical system resource encoding based on feature hashing, the model’s attention to the similarity of system resource access before and after the evolution of malware samples is enhanced. Finally, by designing a contrastive learning strategy, the model’s attention to the similar API fragments retained before and after malware evolution is strengthened. Experimental results show that MME can greatly extend the life-time of the API sequence-based malware detectors and can significantly save the human labeling efforts required for model maintenance. The MME method can be applied to most API sequence-based deep learning malware detection models and help them achieve better sustainable usage.



\bibliographystyle{IEEEtran}
\bibliography{mybibfile}

\vspace{-3em} 
\begin{IEEEbiography}
[{\includegraphics[width=0.8in,height=1.0in,clip,keepaspectratio]{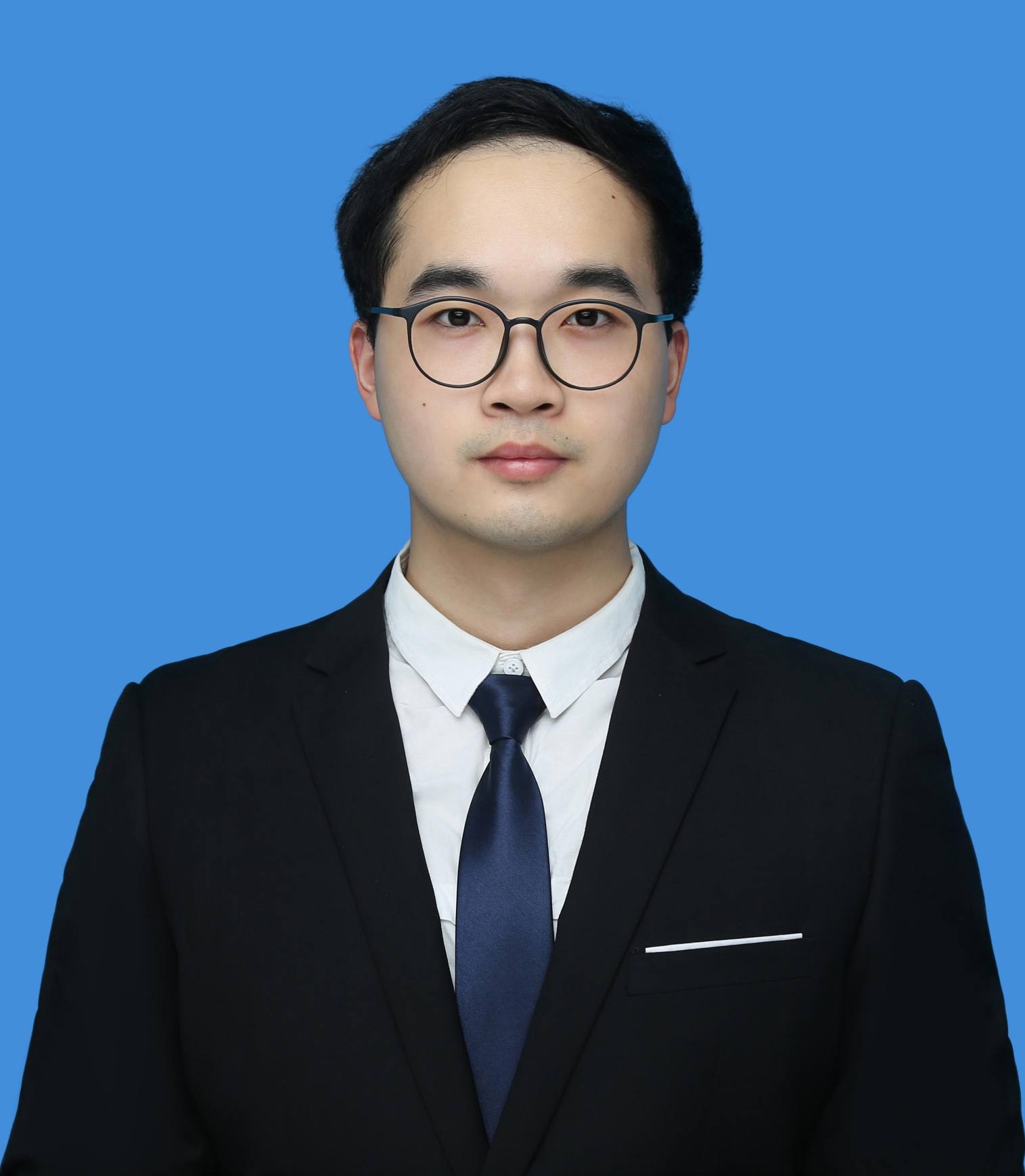}}] 
 Wei Xingyuan is a Ph.D. candidate at the School of Cyber Security, University of Chinese Academy of Sciences, affiliated with the Institute of Information Engineering, Chinese Academy of Sciences. His research focuses on malware detection, AI4Security, LLM4Security, high-security-level cyber defense, and traffic analysis.
\end{IEEEbiography}

\vspace{-5em} 
\begin{IEEEbiography}
[{\includegraphics[width=0.8in,height=1.0in,clip,keepaspectratio]{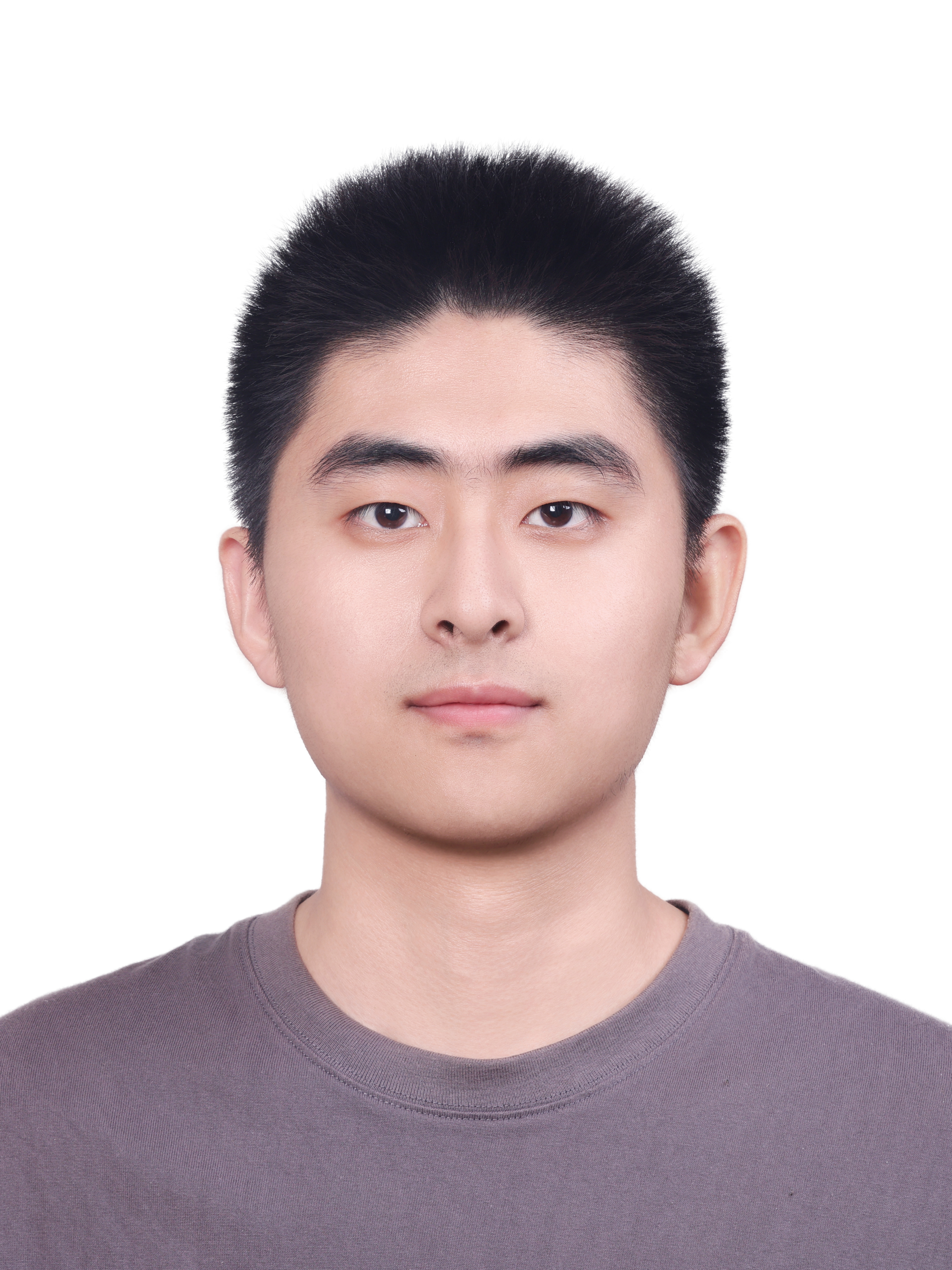}}] 
Dr. Li Ce received his Ph.D. degree from the School of Cyber Security, University of Chinese Academy of Sciences, with his doctoral research conducted at the Institute of Information Engineering, Chinese Academy of Sciences. His research expertise includes malware detection, AI4Security, and LLM algorithms. Currently, he is working at Jiutian Research, China Mobile, Beijing, China.
\end{IEEEbiography}

\vspace{-5em} 
\begin{IEEEbiography}
[{\includegraphics[width=0.8in,height=1.0in,clip,keepaspectratio]{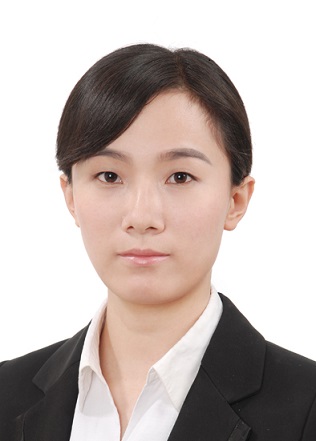}}] 
Dr. Lu QiuJian received his Ph.D. from Beijing University of Posts and Telecommunications. He is currently a Senior Engineer at the Institute of Information Engineering, Chinese Academy of Sciences, and serves as a Master’s Supervisor at the University of Chinese Academy of Sciences. His research focuses on network security monitoring and AI4Security.
\end{IEEEbiography}

\vspace{-5em} 
\begin{IEEEbiography}
[{\includegraphics[width=0.8in,height=1.0in,clip,keepaspectratio]{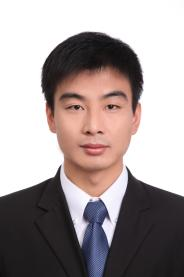}}] 
Li Ning is currently a Senior Engineer at the Institute of Information Engineering, Chinese Academy of Sciences. His research focuses on high-security-level cyber defense, cybersecurity risk assessment, and network security monitoring.
\end{IEEEbiography}

\vspace{-6em} 
\begin{IEEEbiography}
[{\includegraphics[width=0.8in,height=1.0in,clip,keepaspectratio]{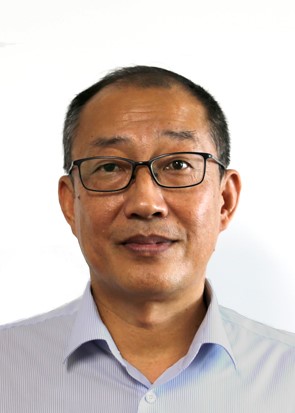}}] 
Sun Degang is a Professor and Doctoral Supervisor at the University of Chinese Academy of Sciences, and holds the title of Professor of Engineering. He has long been engaged in technology R\&D in network and information security and design \& integration of critical information systems.
\end{IEEEbiography}

\vspace{-6em} 
\begin{IEEEbiography}
[{\includegraphics[width=0.8in,height=1.0in,clip,keepaspectratio]{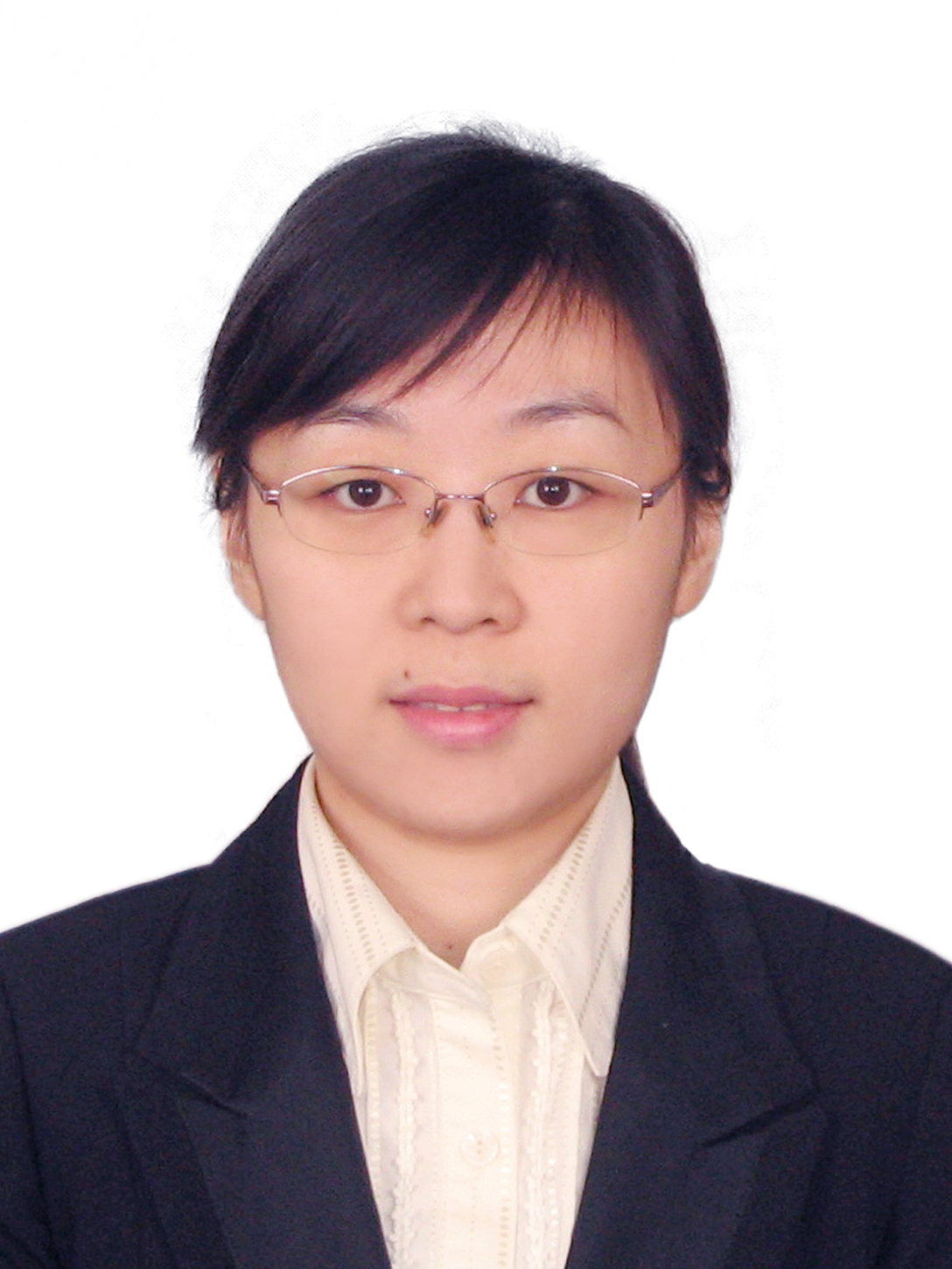}}] 
Wang Yan is a Professor and Master’s Supervisor at the University of Chinese Academy of Sciences, and holds the title of Professor of Engineering. She is currently affiliated with the Institute of Information Engineering, Chinese Academy of Sciences, and has long been dedicated to technology R\&D in network security information systems and high-security-level cyber defense systems development.
\end{IEEEbiography}

\end{document}